%% file: main.tex
\documentclass[a4paper,fleqn]{cas-dc}
\usepackage[authoryear,longnamesfirst]{natbib}
\usepackage[utf8]{inputenc}
\usepackage{amsmath,amssymb,amsfonts}
\usepackage{graphicx}
\usepackage{xcolor}
\usepackage{float}
\usepackage{tikz,pgfplots}
\pgfplotsset{compat=newest}
\usetikzlibrary{shapes.geometric, arrows, arrows.meta, positioning, calc, backgrounds}
\usepackage{nicefrac}
\usepackage{bbm}
\usepackage{url}
\usepackage[export]{adjustbox}

\tikzstyle{midbox} = [rectangle, rounded corners, minimum width=2.62cm, minimum height=.7cm,text centered, draw=black, fill=black!30]
\tikzstyle{arrow} = [thick,->,>=stealth]

\usepackage[breakable,skins,most]{tcolorbox}
\newtcolorbox{summarybox}{
    breakable,
    enhanced,
    width=\linewidth,
    sharp corners=all,
    colback=white!90!black,
    colframe=black,
}
\newtcolorbox{summaryboxOnecolumn}{
    enhanced,
    width=\linewidth,
    sharp corners=all,
    colback=white!90!black,
    colframe=black,
}
\newtcolorbox{supplementbox}{
    breakable,
    enhanced,
    width=\linewidth,
    sharp corners=all,
    colback=white!97!black,
    colframe=black,
    boxrule=0.4pt,
    borderline={0.4pt}{2pt}{black},
}

\DeclareMathOperator*{\argmin}{argmin}

\begin{document}
\let\WriteBookmarks\relax
\def\floatpagepagefraction{1}
\def\textpagefraction{.001}

\shorttitle{An overview of Koopman-based control: From error bounds to closed-loop guarantees}  

\shortauthors{Strässer et al.}  

\title [mode = title]{An overview of Koopman-based control: From error bounds to closed-loop guarantees}  

\tnotemark[1]
\tnotetext[1]{F.\ Allgöwer is thankful that this work was funded by the Deutsche Forschungsgemeinschaft (DFG, German Research Foundation) under Germany's Excellence Strategy -- EXC 2075 -- 390740016 and within grant AL 316/15-1 -- 468094890.
K.\ Worthmann gratefully acknowledges funding by the German Research Foundation (DFG, project ID 535860958) within the research unit ALeSCO --~\textbf{A}ctive \textbf{Le}arning in \textbf{S}ystems and \textbf{Co}ntrol. 
I.\ Mezi{\'c}'s work was partially supported by the Defense Advanced Research Projects Agency (DARPA) under Agreement No. HR00112590152, approved for public release; distribution is unlimited.
R.\ Strässer thanks the Graduate Academy of the SC SimTech for its support.}

\author[1]{Robin Strässer}[orcid=0000-0003-4629-844X]
\cormark[1]
\ead{robin.straesser@ist.uni-stuttgart.de}
\affiliation[1]{organization={University of Stuttgart, Institute for Systems Theory and Automatic Control},
            addressline={Pfaffenwaldring 9}, 
            city={70550 Stuttgart},
            citysep={}, 
            postcode={}, 
            state={Baden-Württemberg},
            country={Germany}}

\author[2]{Karl Worthmann}[orcid=0000-0002-1450-2373]
\ead{karl.worthmann@tu-ilmenau.de}
\affiliation[2]{organization={Technische Universitat Ilmenau, Institute
of Mathematics, Optimization-based Control Group},
            addressline={Weimarer Str. 25}, 
            city={99693 Ilmenau},
            citysep={}, 
            postcode={}, 
            state={Thüringen},
            country={Germany}}

\author[3]{Igor Mezi{\'c}}[orcid=0000-0002-2873-9013]
\ead{mezic@ucsb.edu}
\affiliation[3]{organization={University of California, Santa Barbara, Department of Mechanical Engineering},
            addressline={Engineering II}, 
            city={Santa Barbara},
            postcode={93106}, 
            state={California},
            country={USA}}

\author[1]{Julian Berberich}[orcid=0000-0001-6366-6238]
\ead{julian.berberich@ist.uni-stuttgart.de}

\author[4]{Manuel Schaller}[orcid=0000-0002-8081-5108]
\ead{manuel.schaller@math.tu-chemnitz.de}
\affiliation[4]{organization={Chemnitz University of Technology, Faculty of Mathematics},
            addressline={Reichenhainer Str. 41}, 
            city={09111 Chemnitz},
            citysep={}, 
            postcode={}, 
            state={Sachsen},
            country={Germany}}

\author[1]{Frank Allgöwer}[orcid=0000-0002-3702-3658]
\ead{frank.allgower@ist.uni-stuttgart.de}

\cortext[1]{Corresponding author}

\begin{abstract}
Controlling nonlinear dynamical systems remains a central challenge in a wide range of applications, particularly when accurate first-principle models are unavailable. 
Data-driven approaches offer a promising alternative by designing controllers directly from observed trajectories. 
A wide range of data-driven methods relies on the Koopman-operator framework that enables linear representations of nonlinear dynamics via lifting into higher-dimensional observable spaces. 
Finite-dimensional approximations, such as extended dynamic mode decomposition (EDMD) and its controlled variants, make prediction and feedback control tractable but introduce approximation errors that must be accounted for to provide rigorous closed-loop guarantees.
This survey provides a systematic overview of Koopman-based control, emphasizing the connection between data-driven surrogate models, approximation errors, controller design, and closed-loop guarantees. 
We review theoretical foundations, error bounds, and both linear and bilinear EDMD-based control schemes, highlighting robust strategies that ensure stability and performance. 
Finally, we discuss open challenges and future directions at the interface of operator theory, approximation theory, and nonlinear control.
\end{abstract}


\maketitle

\section{Introduction}\label{sec:introduction}
\noindent
The control of nonlinear dynamical systems remains one of the central challenges in systems and control theory. 
Classical model-based approaches typically rely on accurate first-principle models, which are often difficult to obtain for high-dimensional, uncertain, or complex systems. 
Recent years have therefore seen significant progress in data-driven control methods, which leverage observed trajectories to learn data-driven surrogate models or design controllers directly from data. 
Approaches grounded in robust optimization and the solution of a semi-definite program (SDP), for instance, have established a solid theoretical foundation for data-driven stabilization and performance guarantees across various classes of nonlinear systems~\citep{martin:schon:allgower:2023b,depersis:tesi:2023,berberich:allgower:2024}.

Within this broader landscape, the Koopman operator framework~\citep{koopman:1931,mezic:banaszuk:2000,mezic:banaszuk:2004,mezic:2005} has emerged as an appealing paradigm for the prediction and control of nonlinear dynamical systems.
By lifting nonlinear dynamics into a higher-dimensional space of observables in which the evolution is linear, Koopman-based methods open the door to leveraging the rich toolbox of linear systems theory in inherently nonlinear contexts.
In practice, extended dynamic mode decomposition (EDMD,~\citealp{williams:kevrekidis:rowley:2015}) has become a widely used method for approximating the Koopman operator from data, providing a tractable linear and purely data-driven representation of nonlinear dynamics.
This conceptual shift -- lifting nonlinear dynamics into a higher-dimensional linear space -- has led to a surge of interest~\citep{rowley:mezic:magheri:schlatter:henningson:2009} across diverse application domains, including fluid flows~\citep{bagheri:2013,mezic:2013,taira:hemnati:brunton:sun:duraisamy:bagheri:dawson:yeh:2020}, climate forecasting~\citep{hogg:fonoberova:mezic:2020,azencot:erichson:lin:mahoney:2020}, molecular dynamics~\citep{nuske:klus:2023}, power systems technology~\citep{susuki:mezic:raak:hikihara:2016}, quantum systems~\citep{klus:nuske:peitz:2022}, and even neuroscience~\citep{marrouch:slawinska:giannakis:read:2020,eisen:kozachkov:bastos:donoghue:mahnke:brincat:chandra:tauber:brown:fiete:2024} and cryptography~\citep{schlor:strasser:allgower:2023,strasser:schlor:allgower:2025}. 
Recent advances have further extended the Koopman framework to settings such as parameter-affine systems~\citep{goor:mahony:schaller:worthmann:2025} and networked systems~\citep{schlosser:korda:2022,nandanoori:guan:kundu:pal:agarwal:wu:choudhury:2022,klus:conrad:2024,guo:schaller:worthmann:streif:2025,anantharaman:mauroy:2025}, thereby broadening its reach toward modern large-scale and interconnected dynamical systems.

Naturally, these developments have motivated extensions of Koopman methods to controlled systems. 
Initial formulations~\citep{surana:2016} and subsequent work on model predictive control (MPC)~\citep{korda:mezic:2018a} have demonstrated the feasibility of embedding Koopman models into feedback synthesis. 
From an application perspective, Koopman-based control has shown promise across a wide range of engineering domains, including robotics~\citep{mamakoukas:castano:tan:murphey:2019,bruder:fu:gillespie:remy:vasudevan:2021,haggerty:banks:kamenar:cao:curtis:mazic:hawkes:2023,shi:haseli:mamakoukas:bruder:abraham:murphey:cortes:karydis:2024}, non-holonomic systems~\citep{rosenfeld:kamalapurkar:2024,rosenfelder:bold:eschmann:eberhard:worthmann:ebel:2024}, and human-centered applications such as grip force prediction for robotic rehabilitation~\citep{bazina:kamenar:fonoberova:mezic:2025} and meal detection in glucose monitoring~\citep{tavasoli:shakeri:2025}.
In addition, Koopman operator theory has strong conceptual and methodological links to machine learning. 
Neural ordinary differential equations, for example, have been connected to Koopman-inspired representations~\citep{buzhardt:constante-amores:graham:2025}, while structured approaches such as product Hilbert spaces for control~\citep{lazar:2025} offer principled ways to design controllers with Koopman embeddings.
These developments underscore the increasingly interdisciplinary nature of the field, where control theory, operator-theoretic methods, and data-driven learning converge.

Alongside this growing body of work, several surveys and collections have already mapped parts of the Koopman landscape~\citep{budisic:mohr:mezic:2012,mauroy:mezic:susuki:2020,bevanda:sosnowski:hirche:2021,otto:rowley:2021,brunton:budisic:kaiser:kutz:2022,brunton:zolman:kutz:fasel:2025}. 
Yet, despite the rapid expansion of theory and applications, there remains a need for a unified and rigorous perspective on Koopman-based control, specifically addressing the gap between approximation error bounds in Koopman models and their implications for closed-loop stability and performance guarantees.

This overview aims to fill that gap.
More precisely, we provide a systematic overview of Koopman-based control methods, with a particular focus on the connection between model approximation quality, controller design, and closed-loop guarantees. 
We review the theoretical foundations of the Koopman operator and its finite-dimensional approximations, highlight recent advances in error analysis, and discuss how these insights translate to suitable controller design schemes that guarantee stability and performance for the underlying nonlinear system from data. 
For the latter, we investigate two complementary control approaches, namely feedback design and MPC. 

This article is structured as follows.
In Section~\ref{sec:Koopman}, we provide a concise introduction to the Koopman operator for dynamical systems, with a focus on available error bounds for data-driven approximations.
Section~\ref{sec:EDMDc} reviews the most common \emph{linear} Koopman-based control model, i.e., linear EDMD with control (EDMDc) and related control methods.
In Section~\ref{sec:bilinear:EDMDc}, we discuss extensions of EDMDc to bilinear Koopman models, which enable the derivation of finite-data error bounds.
Section~\ref{sec:controller-design} is devoted to data-driven controller design for nonlinear systems, based on robust control of Koopman approximants together with their error bounds, thereby establishing closed-loop guarantees for the underlying nonlinear dynamics.
Finally, Section~\ref{sec:conclusion} highlights open issues and outlines future research challenges for the community.

\section{Koopman operator}\label{sec:Koopman}
\noindent 
B.O.~Koopman and J.~von Neumann established a connection between nonlinear dynamical systems and linear, but infinite-dimensional ones in their seminal papers~\citet{koopman:1931,koopman:neumann:1932}. 
In this section, we provide a concise introduction to the Koopman operator and its role in data-driven dynamical systems~\citep[cf.][]{mezic:banaszuk:2004,mezic:2005}.
First, we introduce Koopman theory for nonlinear systems in Section~\ref{sec:Koopman:operator}. 
Then, we discuss the spectrum as well as eigenfunctions of the Koopman operator and their role in analysis and prediction in Section~\ref{sec:Koopman:eigenfunctions}.
Section~\ref{sec:Koopman:EDMD} is devoted to EDMD as a tool to generate finite-dimensional and data-driven approximations of the Koopman generator and operator, before finite-data error bounds are discussed in Section~\ref{sec:Koopman:error:bounds}.
A discussion of \emph{kernel EDMD}, a variant of EDMD with data-informed observables, is provided in Section~\ref{sec:Koopman:kernel}.
Finally, Section~\ref{sec:Koopman:further-topics} is devoted to delay embeddings, reprojections, and partial differential equations (PDEs).

\subsection{Koopman operator for dynamical systems}\label{sec:Koopman:operator}

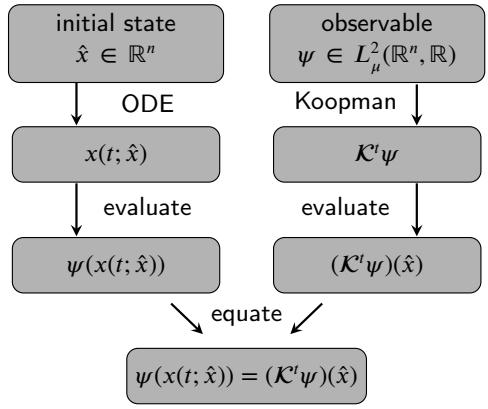
\begin{figure}[tb]
    \centering
    \resizebox{.75\columnwidth}{!}{%
    \begin{tikzpicture}[node distance=1.4cm,shorten <=.25cm, shorten >=.25cm]
	    \node (start) [midbox,text width=2.5cm] {observable \\$\psi\in L^2_\mu(\mathbb{R}^{n},\mathbb{R})$}; 
	    \node (startleft)[left of=start, node distance = .4cm] {};

	    \node (startright)[right of=start, node distance = .5cm] {};	
        \node (startrightarrow)[below of=startright, node distance = .15cm] {};
	    \node (mid) [midbox, below of=start] {$\mathcal{K}^t \psi$};
	    \node (midleft)[left of=mid, node distance = .4cm] {};
	    \node (midright)[right of=mid, node distance = .5cm] {};
	    \draw [arrow] (startrightarrow) --  node[] {} (midright);
	    \node (dummy) [below of = startleft, node distance = .75cm] {\small Koopman};
	    \node (bot) [midbox, below of=mid] {$(\mathcal{K}^t \psi)(\hat{x})$};
	    \node (botleft)[left of=bot, node distance = .4cm] {};
	    \node (botright)[right of=bot, node distance = .5cm] {};
	    \draw [arrow] (midright) -- node[] {} (botright);	    
	    \node (dummy) [below of = midleft, node distance = .65cm] {\small evaluate};
	    \node (start2) [midbox, left of=start,node distance=3.3cm,text width=2.5cm] {initial state\\ $\phantom{L^2_\mu}\hat{x}\in \mathbb{R}^{n}\phantom{L^2_\mu}$}; 
	    \node (start2left)[left of=start2, node distance = .5cm] {};
	    \node (start2leftarrow)[below of=start2left, node distance = .12cm] {};    
         \node (start2right)[right of=start2, node distance = .4cm] {};
	    \node (mid2) [midbox, below of=start2] {$x(t;\hat{x})$};
	    \node (mid2left)[left of=mid2, node distance = .5cm] {};
	    \node (mid2right)[right of=mid2, node distance = .4cm] {};
	    \draw [arrow] (start2leftarrow) -- node[] {} (mid2left);
	    \node (dummy2) [below of = start2right, node distance = .75cm] {\small ODE};
	    \node (bot2) [midbox, below of=mid2] {$\psi(x(t;\hat{x}))$};
        \node (bot2left)[left of=bot2, node distance = .5cm] {};
	    \node (bot2right)[right of=bot2, node distance = .4cm] {};
	    \draw [arrow] (mid2left) -- node[] {} (bot2left);
	    \node (dummy) [below of = mid2right, node distance = .65cm] {\small evaluate};
	    \node (equation) [right of=bot2,node distance = 1.65cm] {};
     \node (end)[midbox,below of=equation]{$\psi(x(t;\hat{x}))=(\mathcal{K}^t\psi)(\hat{x})$};
    \node (asd)[below of=equation,node distance =.65cm]{equate};
    \node(startarrowleft) [below of=botleft, node distance=.2cm] {};
    \node(startarrowright) [below of=bot2right, node distance=.2cm] {}; 
    \draw [arrow] (startarrowleft) -- node[] {} (end);
	\draw [arrow] (startarrowright) -- node[] {} (end);
	\end{tikzpicture}}
    \caption{Schematic sketch of the identity~\eqref{eq:Koopman:identity} (picture modified from~\citet{goor:mahony:schaller:worthmann:2023}).}
    \label{fig:Koopman_identity}
\end{figure}

\noindent We consider dynamical systems governed by time-invariant 
\emph{nonlinear} ordinary differential equations (ODEs)
\begin{equation}\label{eq:ODE}
    \dot{x}(t;\hat{x}) := \frac{\mathrm{d}}{\mathrm{d}t} x(t;\hat{x}) = f(x(t;\hat{x}))     
\end{equation}
with locally Lipschitz-continuous map $f: \mathbb{R}^{n} \rightarrow \mathbb{R}^{n}$ such that local existence and uniqueness of the solution $x(t) = x(t;\hat{x})$ is guaranteed for the initial value problem consisting of~\eqref{eq:ODE} and the initial condition $x(0;\hat{x}) = \hat{x}$.
Tacitly assuming global existence for the time being, Koopman theory allows to trade nonlinearity in exchange for infinite dimensionality based on the identity
\begin{equation}\label{eq:Koopman:identity}
    ( \mathcal{K}^t \psi )(\hat{x}) = \psi( x(t;\hat{x}) ) 
\end{equation}
for all times $t \in [0,\infty)$, initial values $\hat{x} \in \mathbb{R}^{n}$, and observable functions\footnote{An observable (function) is a measurement function of the state. Hence, the output $y = h(x) \in \mathbb{R}^p$ of a dynamical system may be seen as a stacked vector of $p \in \mathbb{N}$ observables.} $\psi \in L^2_\mu(\mathbb{R}^{n},\mathbb{R})$ with finite measure~$\mu$.\footnote{We refer to~\citet[Section~6]{mezic:2020} and the references therein for results on observables on more general function spaces, e.g., tailored to dissipative systems.} 
$(\mathcal{K}^t)_{t \geq 0}$ is the strongly-continuous Koopman semigroup of linear bounded operators; see, e.g.,~\citet[pp.\ 3-33]{mauroy:mezic:susuki:2020} or~\citet[Proposition 3.4]{philipp:schaller:worthmann:peitz:nuske:2025}. 
Identity~\eqref{eq:Koopman:identity} states that one may propagate an observable $\psi \in L^2_\mu(\mathbb{R}^{n},\mathbb{R})$ using the Koopman operator~$\mathcal{K}^t$ and, then, evaluate the propagated observable~$\mathcal{K}^t \psi$ at~$\hat{x}$ instead of evaluating the observable~$\psi$ at $x(t;\hat{x})$, i.e., the solution~$x(\cdot;\hat{x})$ of~\eqref{eq:ODE} at time~$t$; see Figure~\ref{fig:Koopman_identity}.

Alternatively to identity~\eqref{eq:Koopman:identity}, one may invoke the corresponding abstract Cauchy problem $\frac {\mathrm{d}}{\mathrm{d}t} \psi(x(t)) = \mathcal{L} \psi(x(t))$ using the closed and densely-defined Koopman generator~$\mathcal{L}$, which is, in general, an unbounded operator. Using the definition of the infinitesimal generator and~\eqref{eq:ODE} yields
\begin{subequations}\label{eq:Koopman:identity:generator}
    \begin{align}
        \frac{\mathrm{d}}{\mathrm{d}t}\psi(x(t;\hat x)) 
        &= (\mathcal{L} \psi)(x(t;\hat{x})) 
        \\ 
        &= \langle \nabla \psi(x(t;\hat{x})),f(x(t;\hat{x})) \rangle,
    \end{align}    
\end{subequations}
where $\langle \cdot, \cdot \rangle$ denotes the standard scalar product in~$\mathbb{R}^{n}$. 
Equation~\eqref{eq:Koopman:identity:generator} may be seen as the counterpart to the identity~\eqref{eq:Koopman:identity} for the Koopman generator.
Moreover, it shows that one can easily evaluate the Koopman generator if a model of the underlying dynamical system governed by~\eqref{eq:ODE} is available.
Otherwise, such a model has to be estimated from state-derivative data; see, e.g.,~\citet{klus:nuske:peitz:niemann:clementi:schutte:2020}.
Roughly speaking, the relation between the Koopman generator~$\mathcal{L}$ and the semigroup~$(\mathcal{K}^t)_{t \geq 0}$ resembles the relation between a matrix~$A$ and its matrix exponential~$e^{tA}$ for linear time-invariant systems, suitably extended to unbounded operators on Hilbert spaces.

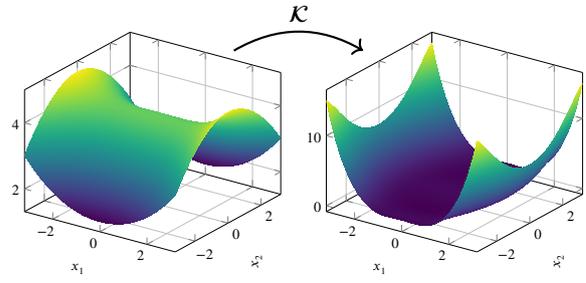
\begin{figure}[tb]
\centering
    \vspace*{0.5\baselineskip}
    \pgfmathdeclarefunction{Fone}{2}{%
      \pgfmathparse{exp(-0.18*(#1*#1 + #2*#2)) * cos(2*#1) * cos(#2)
                    + 0.15*(#1*#1 - #2*#2) + 3}%
    }
    \pgfmathdeclarefunction{Tu}{2}{%
      \pgfmathparse{#1 + 0.12 * #1 * (#1*#1 + #2*#2)}%
    }
    \pgfmathdeclarefunction{Tv}{2}{%
      \pgfmathparse{#2 - 0.12 * #2 * (#1*#1 + #2*#2)}%
    }
    \pgfmathdeclarefunction{Ftwo}{2}{%
      \pgfmathparse{%
        exp(-0.18*( (Tu(#1,#2))^2 + (Tv(#1,#2))^2)) * cos(2*Tu(#1,#2)) * cos(Tv(#1,#2))
        + 0.15*( (Tu(#1,#2))^2 - (Tv(#1,#2))^2 )}%
    }
    \begin{tikzpicture}[scale=.6]
    	\begin{axis}[
            name=left,
            width=7.2cm,height=6.4cm,
            view={35}{30},
            domain=-3.2:3.2, y domain=-3.2:3.2, samples=55,
            axis lines=box, grid=both,
            xlabel={$x_1$}, ylabel={$x_2$},
            colormap/viridis, shader=interp
            ]
            \addplot3[surf]
            ({x},{y},{Fone(x,y)
            }
            );
		\end{axis}
        \begin{axis}[
        at={(left.east)}, anchor=west, xshift=1cm,			width=7.2cm,height=6.4cm,
			view={35}{30},
			domain=-3.2:3.2, y domain=-3.2:3.2, samples=55,
			axis lines=box, grid=both,
			xlabel={$x_1$}, ylabel={$x_2$},
			colormap/viridis, shader=interp
			]
			\addplot3[surf]
			({x},{y},{Ftwo(x,y)
			});
	      \end{axis}
    \end{tikzpicture}
     \begin{tikzpicture}[remember picture, overlay]
        \draw[->, thick, bend left] (-4.7,3.) to node [midway, above] {$\mathcal{K}$} (-3,3.);
    \end{tikzpicture}
    \caption{Illustration of the action of the Koopman operator $\mathcal{K}: \mathcal{C}_b(\Xi) \rightarrow \mathcal{C}_b(\Omega)$, which maps functions to functions, where information is \emph{pulled back} along trajectories.
    Here we use the observable $\varphi(x_1,x_2) = e^{-0.18(x_1^2 + x_2^2)} \cos(2x_1)\cos(x_2) + 0.15(x_1^2 - x_2^2)$ and flow $F(x_1,x_2) = (x_1 (1 + 0.12 \| x \|^2),\, x_2 (1 - 0.12 \| x \|^2))$.}
    \label{fig:Koopman:operator}
\end{figure}

Next, we briefly comment on \emph{discrete-time} nonlinear dynamical systems
\begin{equation}\label{eq:dynamics:DS}
    x^+ = F(x)
\end{equation}
with continuous map $F: \Omega \rightarrow \Xi$ on the open domain $\Omega \subset \mathbb{R}^n$ with Lipschitz boundary~\citep[§4.9]{adams:fournier:2003}\footnote{This Lipschitz condition implies the usual cone conditions for interpolation estimates in reproducing kernel Hilbert spaces; cf.~\citet[Appendix A]{kohne:philipp:schaller:schiela:worthmann:2025} for details.} and $\Xi = F(\Omega) := \{F(x) \mid x \in \Omega\}$.
$F(\hat{x})$ may, e.g., result from the solution of~\eqref{eq:ODE} at time~$\Delta t$ for initial condition~$\hat{x}$, i.e., $F(\hat{x}) = x(\Delta t;\hat{x})$. 
For the discrete-time dynamical system~\eqref{eq:dynamics:DS}, the \emph{linear} Koopman operator~$\mathcal{K}$, which is defined by the identity
\begin{equation}\label{eq:identity:Koopman:DS}
    (\mathcal{K}\varphi)(x) = \varphi(F(x)) 
\end{equation}
for all $\varphi \in \mathcal{C}_b(\Xi)$, $x \in \Omega$,
maps functions in $\mathcal{C}_b(\Xi)$ to $\mathcal{C}_b(\Omega)$, where $\mathcal{C}_b$ denotes the space of bounded continuous functions; see Figure~\ref{fig:Koopman:operator}.\footnote{Alternatively, we could also choose square-integrable observables as in \eqref{eq:Koopman:identity} in case $\Xi = \Omega$ bounded and measurable and with measure~$\mu$ induced by the normalized characteristic function~$\chi_\Xi$ as density.}

The restriction of the map~$F$ to the set~$\Omega$ highlights that the Koopman operator actually maps functions defined on $\Xi$ to functions defined on the domain~$\Omega$ of~$F$, which explains the terminology \emph{pullback operator}~\citep[cf.][]{avila:mezic:2023} and is helpful to understand the Koopman operator as the adjoint of the Perron--Frobenius operator, which propagates densities \emph{forward in time} and is, thus, a \emph{push-forward operator}.
\begin{supplementbox}
    \emph{Supplementary material}: \textbf{Koopman operator is of transport type}.
    
    Recalling the abstract Cauchy problem $\frac {\mathrm{d}}{\mathrm{d}t} \psi(x(t)) = \mathcal{L} \psi(x(t))$ and plugging in the identity~\eqref{eq:Koopman:identity:generator} yields the transport equation
    \begin{equation*}
        \frac {\mathrm{d}}{\mathrm{d}t}  \psi(x(t)) = \langle \nabla \psi(x(t)), f(x(t)) \rangle,
    \end{equation*}
    i.e., a first-order hyperbolic PDE. Further, it may be used to approximate conservation laws from data; see~\citet{kaiser:kutz:brunton:2018} using Lie and Poisson brackets and the physics-informed EDMD (piEDMD,~\citealp{baddoo:herrmann:mckeon:kutz:brunton:2023}), going beyond conservation, but also learning features such as causality, shift-invariance, and locality, in particular for PDEs.

    The \textbf{Perron--Frobenius operator} is the (formal) adjoint of the Koopman operator, hence also a transport-type operator. It propagates probability densities $\rho \in L^1(\Omega)$ via 
    \begin{align*}
        \mathcal{P}\rho(x) = \int_{\Omega} p(y,x)\rho(y)\,\mathrm{d}y,
    \end{align*}
    where, for general stochastic systems, $p(y,x)$ is the transition kernel modeling the conditional probability that $F(y)=x$. For deterministic systems, we have $p(y,x) = \delta_{x}(\Phi(y))$.\footnote{For simplicity, we have imposed forward invariance of the set~$\Omega$ ensuring that the successor distribution is, again, a probability distribution.}
    For further reading, we point the reader to the overview article~\citet{klus:nuske:koltai:wu:kevrekidis:schutte:noe:2018}.
    Recent applications of the Perron--Frobenius operator in deep learning using reproducing kernel Hilbert $C^*$-modules can be found in~\citet{hashimoto:ikeda:kadri:2023}.
\end{supplementbox}

\subsection{Eigenfunctions}\label{sec:Koopman:eigenfunctions}
\noindent
A variety of papers build upon representations leveraging Koopman-invariant subspaces, e.g., for controller design~\citep{korda:mezic:2020}. 
A straightforward way to define a Koopman-invariant subspace is to consider eigenfunctions of the Koopman operator; see, e.g., \citet{mezic:banaszuk:2004,mezic:2005,budisic:mohr:mezic:2012,mauroy:mezic:2012,brunton:brunton:proctor:kutz:2016,mezic:2021}.
Here, \citet{mezic:2021} provides a proof that such linear representations exist if and only if the point spectrum is non-empty.

For the discrete-time dynamical system~\eqref{eq:dynamics:DS} (or the Koopman generator~$\mathcal{L}$ as well as the Koopman operator~$\mathcal{K}^{\Delta t}$ for a given time step~$\Delta t > 0$) a Koopman eigenfunction $\psi \in L^2_\mu(\mathbb{R}^n,\mathbb{C})$, $\psi \not \equiv 0$, for a corresponding eigenvalue~$\lambda \in \mathbb{C}$, satisfies 
\begin{equation}\label{eq:Koopman:eigenfunction}
      \psi(x^+) 
      \stackrel{\eqref{eq:dynamics:DS}}{=} \psi(F(x)) \stackrel{\eqref{eq:identity:Koopman:DS}}{=} \mathcal{K} \psi(x) 
      = \lambda \psi(x);
\end{equation}
see also the identity~\eqref{eq:Koopman:identity}.\footnote{In this subsection, we use w.l.o.g.\ complex-valued observables to emphasize their utility as eigenfunctions due to their clear geometric meaning; see, e.g., \citet{mezic:2005} and \citet{rowley:mezic:magheri:schlatter:henningson:2009}.}
Hence, if we restrict the Koopman operator to the finite-dimensional subspace
\begin{equation*}
    \mathbb{V} = \operatorname{span} \{\psi_1,\ldots,\psi_M\}, 
\end{equation*}
where $\psi_i$ are eigenfunctions with eigenvalues $\lambda_i$, $i \in \{1,...,M\}$, we have for observables $\psi \in \mathbb{V}$ the implication
\begin{equation}\label{eq:eigenfunction_propagated}
   \psi = \sum_{i=1}^M c_i \psi_i\quad \Longrightarrow \quad \psi(x^j) = \sum_{i=1}^M c_i \lambda_i^j \psi_i.
\end{equation}
In this expansion, $x^j$ stands for the state resulting from applying the dynamics~\eqref{eq:dynamics:DS} $j\in\mathbb{N}$ times.
Further, $c_i$, $i \in \{1,...,M\}$, are the \emph{Koopman modes} of the observable $\psi$; see~\citet{mezic:2005,rowley:mezic:magheri:schlatter:henningson:2009} and, e.g., \citet{mezic:2013,bagheri:2013} for an analysis of Koopman modes in fluid dynamics. 
In this decomposition, each mode represents a distinct spatial pattern evolving at a characteristic temporal frequency.
In matrix notation, and using the vector-valued observable defined by
\begin{equation}\label{eq:observable:vectorized}
    \Psi(x) = \begin{pmatrix}
        \psi_1(x) & \cdots & \psi_M(x)
    \end{pmatrix}^\top
\end{equation}
with $M \in \mathbb{N}$,~\eqref{eq:eigenfunction_propagated} reads $\Psi(x^+) = \Lambda \Psi(x)$, where $\Lambda = \operatorname{diag}(\lambda_1,\ldots,\lambda_M) \in \mathbb{C}^{M\times M}$. 

Representations and expansions in terms of eigenfunctions are particularly appealing due to, e.g., their value for analysis~\citep{mezic:2015,mauroy:mezic:2016} and long-term predictions; see~\citet{korda:putinar:mezic:2020} and~\citet[Figure~13]{giannakis:valva:2024}, where predictions are made on timescales orders of magnitude larger than the ones given by the Lyapunov exponents.
Moreover, if the dictionary~$\mathbb{V}$ forms a Koopman-invariant (finite-dimensional) subspace, i.e., $\mathcal{K}\psi \in \mathbb{V}$ for all $\psi \in \mathbb{V}$, then no projection onto that subspace is required to describe the dynamical behavior as long as the observable can be represented as a linear combination of~$\psi_1,\ldots,\psi_M$, i.e., it is contained in~$\mathbb{V}$. 
More precisely, this means that we have an exact finite-dimensional representation of the action of the Koopman operator on~$\mathbb{V}$.
Then, eigenfunctions can be employed to conduct a stability analysis of (hyperbolic) fixed points and limit cycles~\citep{mauroy:mezic:2016}. 
Consequently, there has been a lot of research towards the discovery of informative Koopman-invariant subspaces~\citep{pan:arnold-medabalimi:duraisamy:2021} and their optimal construction~\citep{korda:mezic:2020}. 
Further,~\citet{mezic:2020,korda:mezic:2020,liu:ozay:sontag:2025} point out fundamental links to non-recurrent sets.

While the above works showcase that eigenfunctions are of fundamental use in prediction and control, the corresponding analysis and computation (in particular by data-driven methods) remain challenging. 
The reason for this is the infinite-dimensional nature of the Koopman operator, such that the structure of the spectrum~$\sigma(\mathcal{K})$, i.e., the complement of the resolvent set in the complex plane, i.e., 
\begin{equation*}
    \sigma(\mathcal{K}) := \mathbb{C} \setminus \{ \lambda \in \mathbb{C} \mid (\lambda I - \mathcal{K})^{-1}\in L(L^2_\mu(\mathbb{R}^n,\mathbb{C})) \}, 
\end{equation*}
may be far more intricate than in finite dimensions. 
It contains eigenvalues as defined by~\eqref{eq:Koopman:eigenfunction}, the point spectrum $\sigma_p(\mathcal{K})$, and --~particularly for Koopman operators~-- the continuous spectrum $\sigma_c(\mathcal{K})$ describing, e.g., irreducible nonlinear dynamics such as chaos~\citep{mezic:2005}; see, e.g.,~\citet[Section 2.1.2]{colbrook:2024} for further details. 
The latter are essential in Koopman-based modeling and appear in a wide range of important applications.

\subsection{Extended dynamic mode decomposition}\label{sec:Koopman:EDMD}
\noindent
To streamline the presentation, we now present EDMD as a data-driven numerical approximation of the Koopman operator, leveraging samples from a compact set~$\mathbb{X} \subset \mathbb{R}^{n}$ for a given and, thus, fixed time step $\Delta t > 0$.%
\footnote{For a flexible framework for general samples drawn from a finite Borel measure $\mu$, or sampling along sufficiently long trajectories if the system is ergodic~\citep[see][Chapter 5]{lasota:mackey:2013}, we refer to~\citet{philipp:schaller:worthmann:peitz:nuske:2025}.}
To compute an approximation of the Koopman operator~$\mathcal{K}^{\Delta t}$ on~$\mathbb{X}$, we first choose \emph{finitely-many} (linear independent) observables, i.e., functions~$\psi_1,\ldots,\psi_M \in L^2_\mu(\mathbb{X},\mathbb{R})$, which compose the dictionary~$\mathbb{V} = \operatorname{span}\{ \psi_1,\ldots,\psi_M \}$.
Possible choices include, e.g., monomials of degree less or equal $n_d \in \mathbb{N}$ resulting in $M = \sum_{k=0}^{n_d} \binom{k+n-1}{n-1}$ observables. 
Then, using \emph{finitely-many} data points 
\begin{equation*}
    \mathcal{X} := \{x_1,\ldots,x_d\} \subset \mathbb{X} \subset \mathbb{R}^{n}
\end{equation*}
with their respective successors $x(\Delta t;x_k)$, $k \in \{1,...,d\}$, the linear regression problem\footnote{Note that we identified the respective operator with its matrix representation on~$\mathbb{V}$.}
\begin{equation}\label{eq:EDMD:regression-problem}
    \min_{K \in \mathbb{R}^{M \times M}} \| \Psi_Y - K \Psi_{X} \|_F
\end{equation}
is solved in EDMD~\citep{tu:2013} with $(M \times d)$-data matrices
\begin{subequations}\label{eq:Koopman:EDMD:lifted-data-matrices}
    \begin{align}
        \Psi_X &:= \begin{pmatrix}
            \Psi(x_1) & \cdots & \Psi(x_d)
        \end{pmatrix} \quad\text{ and }\\ 
        \Psi_Y &:= \begin{pmatrix}
            \Psi(x(\Delta t;x_1)) & \cdots & \Psi(x(\Delta t;x_d))
        \end{pmatrix},
    \end{align}
\end{subequations}
where the subscript~$F$ indicates the Frobenius norm.
Here, we use the successor data in $\Psi_Y$ since the map $f$ is unknown.
The transpose $K^\top$ of the solution $K = K_d^M = (\Psi_Y \Psi_X^\top) (\Psi_X \Psi_X^\top)^{-1}$ of the regression problem~\eqref{eq:EDMD:regression-problem} may be considered as a finite-dimensional linear approximation of the Koopman operator~$\mathcal{K}^{\Delta t}$; see~\citet{williams:kevrekidis:rowley:2015,korda:mezic:2018b,mezic:2022} for details. 
Then, with the concatenated vector of observables~$\Psi$ defined by~\eqref{eq:observable:vectorized}, the Koopman action may be approximated via application of the EDMD estimator, i.e.,
\begin{equation*}
    \Psi(x(\Delta t;\hat{x})) = (\mathcal{K}^{\Delta t} \Psi)(x) \approx K \Psi(x),
\end{equation*}
where the Koopman operator is applied element-wise to each element in $\Psi$.
To be more precise, $K$ is an empirical estimate of the compression $P_{\mathbb{V}} \mathcal{K}^{\Delta t}|_{\mathbb{V}}$ (or Galerkin projection) of the Koopman operator~$\mathcal{K}^{\Delta t}$, i.e., the projection on the finite-dimensional subspace~$\mathbb{V}$ of the Koopman operator restricted to~$\mathbb{V}$.

If the dictionary is spanned by a set of eigenfunctions, the projection~$P_{\mathbb{V}}$ is not needed, which renders this case of particular interest~\citep{brunton:brunton:proctor:kutz:2016}; see Section~\ref{sec:Koopman:eigenfunctions}. 
This has motivated the in-depth study of Koopman invariant subspaces~\citep{takeishi:kawahara:yairi:2017}, which leads to advanced concepts like the consistency index~\citep{haseli:cortes:2023a} to balance expressiveness and accuracy; see, e.g.,~\citet{haseli:cortes:2023b} and the references therein. 
To this end, dictionary-learning techniques have been developed~\citep{li:dietrich:bollt:kevrekidis:2017,lusch:kutz:brunton:2018} ---~including approaches based on analytical constructions~\citep{shi:karydis:2021}, Koopman eigenfunction kernels~\citep{bevanda:beier:lederer:sosnowski:hullermeier:hirche:2023}, and deep neural networks~\citep{yeung:kundu:hodas:2019}.
Further,~\citet{korda:mezic:2020} provide an extension to control systems and~\citet{bevanda:beier:kerz:lederer:sosnowski:hirche:2022} solve dictionary-learning regression and reconstruction at once for stable systems.
Moreover, prior knowledge can be incorporated in EDMD, e.g., in piEDMD~\citep{baddoo:herrmann:mckeon:kutz:brunton:2023}.
However, since the Koopman operator has, in general, also a continuous spectrum (compare Section~\ref{sec:Koopman:eigenfunctions}), more sophisticated techniques are needed to analyze spectral properties of the Koopman operator in a rigorous manner~\citep{korda:putinar:mezic:2020,colbrook:ayton:szoke:mate:2023}.
For dynamic mode decomposition (DMD) variants with guarantees in spectral approximation, we mention the residual DMD (ResDMD,~\citealp{colbrook:townsend:2024}) or rigged DMD, particularly suited for approximation of the resolvent and continuous spectra~\citep{colbrook:drystrale:horning:2025}. 
For an introduction to this topic, we refer to~\citet[Section 2]{colbrook:2024}, where we point particularly to the aspect of pseudospectra measuring the stability of the spectrum under perturbations that are inevitably present in finite-data and finite-dictionary computations. 
Sharp rates in data-driven spectral computations are provided in~\citet{kostic:lounici:novelli:pontil:2023} and the recently-proposed framework in~\citet{colbrook:mezic:stepanenko:2024}, revealing fundamental barriers in (spectral) approximability.
A perspective using Liouville operators is given in~\citet{rosenfeld:kamalapurkar:2023}, introducing singular DMD achieving convergence by compactness of the Koopman generator by suitably adapting the spaces to provide well-defined dynamic modes.
Last, we mention that the numerical stability of \eqref{eq:EDMD:regression-problem} may be improved, e.g., by approximating the pseudoinverse in the definition of the Koopman surrogate by means of a truncated singular value decomposition as suggested already in \citet{williams:hemati:dawson:kevrekidis:rowley:2016} or adding a regularization such as sparsity-promoting terms~\citep{hou:sanjari:2023}.

\subsection{Finite-data error bounds}\label{sec:Koopman:error:bounds}
\noindent
Recently, various error bounds on the EDMD approximant have been derived. 
The respective approximation error results from the projection on~$\mathbb{V}$ (finitely many, i.e., $M$~observables) and the estimation (finitely many, i.e., $d$~data points).
In the infinite-dictionary ($M \rightarrow \infty$) and infinite-data limit ($d\to\infty$),~\citet{korda:mezic:2018b} show that EDMD converges strongly to the true Koopman operator.
The first probabilistic finite-data bounds on the \emph{estimation error} are given in~\citet{mezic:2022} for deterministic ergodic systems, assuming sampling on a sufficiently long trajectory under the assumption that the spectrum of the Koopman operator is discrete and non-dense on the unit circle. 
First results on both the estimation and approximation error for i.i.d.\ (independently and identically distributed) sampling of deterministic systems can be found in~\citet{zhang:zuazua:2023}, leveraging, among others, finite-element techniques originally developed for the numerical approximation of hyperbolic PDEs. 
For systems governed by stochastic differential equations with Wiener process~$W_t$\footnote{For a more general setting using random dynamical systems, we refer to~\citet{wanner:mezic:2022,vcrnjaric:macevic:mezic:2020}.}, i.e., 
\begin{equation}\label{eq:SDE}
    \mathrm{d}X_t = f(X_t)\,\mathrm{d}t + \sigma(X_t)\, \mathrm{d}W_t,
\end{equation}
the Koopman operator is defined via the conditional expectation $\mathcal{K}^{\Delta t}\psi(\hat{x}) = \mathbb{E}(\psi(X_{\Delta t}) \mid X_0 = \hat{x})$.
\citet{nuske:peitz:philipp:schaller:worthmann:2023} derive first finite-data error bounds of the form
\begin{equation*}
    \mathbb{P}( \| K_d^M - P_{\mathbb{V}} \mathcal{K}^{\Delta t}|_{\mathbb{V}} \|_F \leq \varepsilon ) \geq 1 - \delta
\end{equation*}
for given accuracy~$\varepsilon$, probabilistic tolerance~$\delta$, and i.i.d.\ or ergodic sampling\footnote{Ergodic sampling corresponds to sampling along sufficiently long trajectories for ergodic systems; see~\citet{lasota:mackey:2013}. 
Alternatively, one may enforce ergodicity by suitable sampling strategies as proposed, e.g., by~\citet{miller:murphey:2013} and~\citet{miller:silverman:maciver:murphey:2015} for control systems.}, including bounds on the generator and the first extension to control systems. 
The required amount of data~$d$ depends quadratically on $\varepsilon^{-1}$ and~$M$, i.e., the dimension of the ansatz space~$\mathbb{V}$, as well as linearly on~$\delta^{-1}$.
Hence, the required amount of data is $d = \mathcal{O}(M^2/(\delta \varepsilon^2))$.
However, the result for ergodic sampling hinges on the exponential stability of the Koopman semigroup.
\citet{philipp:schaller:boshoff:peitz:nuske:worthmann:2024} deduce bounds on the estimation error under non-restrictive assumptions for deterministic and stochastic dynamical systems (on Polish spaces) in discrete and continuous time, leveraging novel variance representations.
However, all error bounds have in common that the full approximation error is only bounded in a least-squares ($L^2$) sense~\citep{zhang:zuazua:2023,schaller:worthmann:philipp:peitz:nuske:2023} in view of the employed dictionary of finite elements to cope with the projection error ---~if the latter is analyzed at all. 
The underlying reasoning is that the Koopman operator corresponds to a transfer operator and the respective dynamics are, thus, governed by a hyperbolic PDE; see Section~\ref{sec:Koopman:operator}.
For a recent work providing results for noisy data, we refer to~\citet{llamazares:llamazares:latz:klus:2024}, where convergence in the infinite-dictionary and infinite-data limit is given, but no convergence rates (and thus in particular no finite-data error bounds) are specified.
\citet{bevanda:beier:capone:sosnowski:hirche:lederer:2025} further proposed a method using a trajectory-based \emph{Koopman equivariance} in order to learn new probabilistic Koopman representations with enhanced generalization capacities to tractably track the accompanying uncertainties.

\subsection{Kernel EDMD}\label{sec:Koopman:kernel}
\noindent
Kernel EDMD (kEDMD; see, e.g.,~\citealp{klus:nuske:hamzi:2020}) is particularly appealing since the dictionary is chosen in a data-based manner. 
More precisely, the elements~$\psi_j$, $j \in \{1,...,M\}$, spanning the dictionary are given by the canonical features, i.e., the kernel~$k(x_j,\cdot)$ of the native space (reproducing kernel Hilbert space; RKHS) evaluated at the data points~$x_j$. 
Hereby, already the terminology indicates one of the key features, the reproducing property
\begin{equation}\label{eq:property:reproducing}
    f(x) = \langle f,\phi_x \rangle_{\mathbb{H}} \qquad\forall\,x \in \mathbb{R}^n,
\end{equation}
where $\phi_x = k(x,\cdot)$.
On the one hand, the property~\eqref{eq:property:reproducing} allows for efficient evaluation of inner products by function evaluations. 
On the other hand, point evaluation is well defined, which is important to derive uniform error bounds ($L^\infty$); see, e.g.,~\citet{philipp:schaller:worthmann:peitz:nuske:2024} for first bounds on the estimation error of kEDMD. 
Nevertheless, it is worth noting that in~$L^2$, ergodic theorems or the central limit theorem allow for deriving error bounds on kEDMD without pointwise evaluation; see~\citet{mezic:2022}.

Herein, suitable choices of the kernel~$k$ are key. 
For example,~\citet{gonzales:abudia:jury:kamalapurkar:rosenfeld:2023} point out that the Koopman operator is only bounded on the RKHS of the Gaussian radial basis function (RBF) kernel if the dynamics are affine. 
For a remedy for constructing suitable invariant subspaces for dissipative systems in Fock spaces, we refer to~\citet{mezic:2020}.
Other widely used kernels are Wendland functions~\citep{wendland:2004}, particular compactly supported RBFs with smoothness degree~$s$, for which the RKHS $\mathbb{H} = \mathcal{N}_{\Phi_{n,s}}(\Omega)$ has been shown to be isomorphic to Sobolev spaces of weakly differentiable functions, that is $\mathcal{N}_{\Phi_{n,s}} \cong H^{\sigma_{n,s}}$, $\sigma_{n,s} := \tfrac {n+1}{2} + s$.\footnote{Matérn kernels, which are also radially symmetric, are a valid alternative giving rise to the same native spaces.} 
To analyze kEDMD for Wendland kernels,~\citet[Section~4]{kohne:philipp:schaller:schiela:worthmann:2025} show that, supposing that the flow map is $C^m$, $m>n/2$, Sobolev regularity is preserved.
This means that we have $\mathcal{K}(\mathcal{N}(\Xi)) \subseteq \mathcal{N}(\Omega)$, i.e., Koopman invariance of the RKHS and, thus, boundedness of the Koopman operator $\mathcal{K}: \mathcal{N}(\Xi) \rightarrow \mathcal{N}(\Omega)$, i.e., $\| \mathcal{K} \|_{\mathcal{N}(\Xi) \rightarrow \mathcal{N}(\Omega)} < \infty$ for general nonlinear dynamical systems~\eqref{eq:dynamics:DS}.\footnote{We point out that there may exist non-smooth $L^2$-eigenfunctions~\citep{mezic:banaszuk:2004,mezic:2005}, explaining the large body of literature on eigenfunction approximation using a Koopman operator defined on~$L^2$.}

\begin{figure}[tb]
    \centering
    \includegraphics[width=1.\columnwidth]{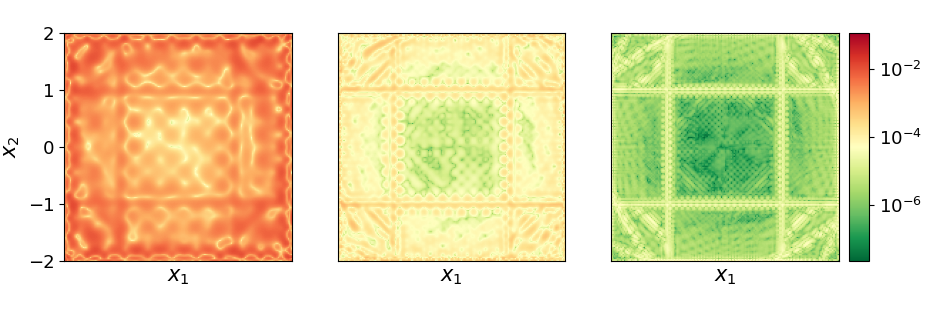}
    \caption{Approximation error for a propagated coordinate map with decreasing fill distance~$h_\mathcal{X}$ (left to right) using a Padua grid. Figure modified from~\citet[Figure~1]{bold:philipp:schaller:worthmann:2024}.}
    \label{fig:kEDMD:error-Padua}
\end{figure}

This invariance result is essential to establish uniform ($L^\infty$) error bounds for kEDMD on the approximation error for deterministic dynamical systems in~\citet{kohne:philipp:schaller:schiela:worthmann:2025}; see also~\citet{kurdila:paruchuri:powell:guo:bobade:estes:wang:2024} for preliminary results.
In kEDMD, the propagation of an observable $\psi$ via the Koopman operator corresponding to any autonomous (discrete-time) dynamics~\eqref{eq:dynamics:DS}, i.e., $x^+ = F(x)$, can be approximated by the kernel-based estimator 
\begin{equation}\label{eq:bilinear:propagation-observable}
    \psi(x^+) = (\mathcal{K}\psi)(x) 
    \approx \sum\nolimits_{j=1}^d (\hat{K} \psi_{\mathcal{X}})_j\phi_{x_j}(x)
\end{equation}
with $\hat{K} = K_{\mathcal{X}}^{-1} K_{F(\mathcal{X})}K_{\mathcal{X}}^{-1}$,
$\psi_{\mathcal{X}} = \begin{pmatrix}\psi(x_1) & \cdots & \psi(x_d) \end{pmatrix}^\top$, and
\begin{align*}
        K_{F(\mathcal{X})} 
        &= \begin{pmatrix}
            \mathsf{k}(x_1,F(x_1)) & \cdots & \mathsf{k}(x_d,F(x_1)) \\
            \vdots & \ddots & \vdots \\
            \mathsf{k}(x_1,F(x_d)) & \cdots & \mathsf{k}(x_d,F(x_d))
        \end{pmatrix}.
\end{align*}
The key observation is that kEDMD yields the approximant $\widehat{\mathcal{K}}= P_\mathcal{X} \mathcal{K}P_\mathcal{Y}$, where $P_\mathcal{X}$ is the $\mathbb{H}$-orthogonal projection solving the regression problem~\eqref{eq:EDMD:regression-problem}. Then, we may compute
\begin{eqnarray*}
    \| \widehat{\mathcal{K}} - \mathcal{K} \|_{\mathbb{H}\to L^\infty} & = & \| P_{\mathcal{X}} \mathcal{K} ( P_{\mathcal{Y}} - I) + (P_{\mathcal{X}} - I) \mathcal{K} \|_{\mathbb{H}\to L^\infty} \\
    & \leq & C_1 h_{\mathcal{X}}^{k+\frac 12} + C_2 h_{\mathcal{Y}}^{k+\frac 12}
\end{eqnarray*}
with fill distances
\begin{equation*}
    h_\mathcal{X} := \max_{x_i, x_j \in \mathcal{X}} \| x_i - x_j \| \quad\text{ and }\quad h_\mathcal{Y} := \max_{y_i, y_j \in \mathcal{Y}} \| y_i - y_j \|
\end{equation*}
with $\mathcal{Y} := \{ F(x_1),\ldots,F(x_d) \}$; see Figure~\ref{fig:kEDMD:error-Padua}.
Therein, $\| P_{\mathcal{X}} \|_{\mathcal{N}(X) \rightarrow \mathcal{N}(X)} = 1$ (projection), $\| \mathcal{K} \|_{L^\infty \rightarrow L^\infty} = 1$, and approximation-theoretic results are leveraged to bound $\| P_{\mathcal{X}}f - f \|_{\mathcal{N}(\Omega) \rightarrow L^\infty}$ and $\| P_{\mathcal{Y}}f - f \|_{\mathcal{N}(\Xi) \rightarrow L^\infty}$ by means of the fill distance, respectively.
We emphasize that this approximates the (infinite-dimensional) Koopman operator, meaning that the estimate is directly applicable for arbitrary observable functions, e.g., outputs of particular interest.

An appealing option to reduce complexity in kernel-based methods, such as kEDMD, are random Fourier features~\citep{nuske:klus:2023} or the Nyström method~\citep{caldarelli:chatalic:colome:molinari:ocampo-martinez:torras:rosasco:2024}, also called sketching~\citep[Section~5]{bevanda:driessen:iacob:toth:sosnowski:hirche:2024}; see also~\citet{meanti:chatalic:kostic:novelli:pontil:rosasco:2024} for error bounds.
Beyond that,~\citet{yadav:mauroy:2025} propose an alternative based on Bernstein polynomials in case the data is given in a lattice-like structure.
Moreover,~\citet{kostic:novelli:maurer:ciliberto:rosasco:2022} analyze the setting of sampling along trajectories in an RKHS, embedding kEDMD into the broader picture of statistical learning and providing error bounds in terms of mixing conditions ensuring ergodicity. 
Last, we refer to~\citet{boulle:colbrook:conradie:2025} for a convergence analysis of decompositions and spectral approximations in RKHS as discussed in Subsection~\ref{sec:Koopman:eigenfunctions}.

\subsection{Delay embeddings, reprojections, and PDEs}\label{sec:Koopman:further-topics}
\noindent
While prediction in the lifted space is appealing due to linear dynamics, data is typically only available from observables corresponding to (components of) the output~$y = h(x)$. 
The output can either represent sensor measurements in real-world applications or other physically meaningful functions.
This observation might severely restrict the freedom w.r.t.\ the choice of the dictionary~$\Psi$ and, thus, complicate the construction of, e.g., Koopman-invariant subspaces or even the applicability of EDMD. 
A potential remedy is the use of delay coordinates corresponding to measurement sequences as first proposed in~\citet{mezic:banaszuk:2004} and we refer to~\citet{susuki:mezic:2015,arbabi:mezic:2017b} for the required changes in DMD. 
Then, results on delay embeddings using Takens' theorem~\citep{noakes:1991} have been invoked, e.g., in~\citet{robinson:2005}, for the theoretical analysis, and we refer to~\citet{mezic:2022,koltai:kunde:2024} for an extension of the error analysis.
\citet{kamb:kaiser:brunton:kutz:2020} have successfully incorporated delay embeddings, leading to the Hankel (Alternative) View of Koopman (HVOK/HAVOK).
This approach has been further studied in~\citet{pan:duraisamy:2020} and it has produced convincing results, e.g., in flow prediction~\citep{yuan:zhou:zhou:wen:liu:2021}.
Notably, follow-up work encompasses the latent EDMD framework~\citep{ouala:chapron:collard:gaultier:fablet:2023} and uses, in addition, deep delay autoencoders in~\citet{bakarji:champion:kutz:brunton:2023}; see also~\citet{peitz:harder:nueske:philipp:schaller:worthmann:2025} for first results pointing to PDEs and~\citet{otto:peitz:rowley:2024} for an extension to control.

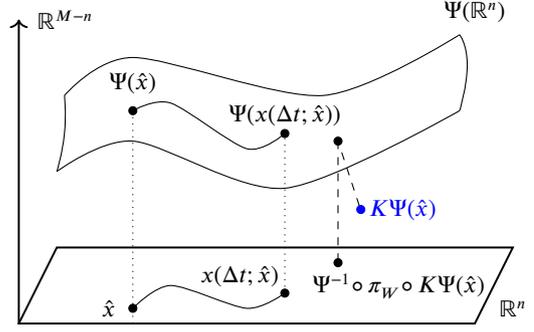
\begin{figure}[tb]
    \centering
    \input{figures/reprojection}
    \caption{The lifted dynamics evolve on a \emph{nonlinear} manifold, which may not be invariant w.r.t.\ the EDMD surrogate such that reprojections $\pi_W:\mathbb{R}^{M}\to \operatorname{im}\Psi$ may be required to preserve consistency. Figure taken from~\citet{goor:mahony:schaller:worthmann:2023}.}
    \label{fig:EDMD:reprojection}
\end{figure}

Another important aspect is the reinterpretation of the lifted state in the original state space of the dynamical system~\eqref{eq:dynamics:DS}. 
To this end,~\citet{mauroy:2021,peitz:otto:rowley:2020} and others suggest to include the coordinate functions in the dictionary. 
This allows one to \emph{project back} from the lifted to the original space and, then, potentially lift again to ensure consistency in the lifted space.
For example, if the observables $\psi_1(x) = x_1$ and $\psi_2(x) = x_1^2$ are contained in~$\mathbb{V}$, the relation $\psi_2(x) = \psi_1(x)^2$ is preserved by the true Koopman operator, but might be violated by the propagation forward in time using the data-driven surrogate model based on EDMD; see Figure~\ref{fig:EDMD:reprojection}.
However, using the so-called coordinate reprojection, one may reproject data-driven predictions to the manifold of consistent states. 
For more elaborated reprojection techniques leveraging differential-geometric arguments and allowing a maximum-likelihood interpretation by inducing a weighted reprojection $\pi_W:\mathbb{R}^M\to \operatorname{im}(\Psi)$, we refer to~\citet{goor:mahony:schaller:worthmann:2025}. 
Herein, also the necessity to include the coordinate maps and, thus, essentially the whole state in~$\Psi$ is removed. 
Rather, one may use the detectability concept, e.g., based on recent results in~\citet{zhang:pan:scattolini:yu:xu:2022} using existence results inferred from the inverse function theorem~\citep{clarke:1976} or~\citet{iacob:szecsi:mate:beintema:schoukens:toth:2025}, where the latter nicely fits to the likelihood framework.

A problem that is yet far from being solved concerns finite-data error bounds for Koopman-based methods for PDEs, which are crucial as a foundation for data-driven control of distributed parameter systems. 
For autonomous systems, there are several works exploring Koopman theory for PDEs; see~\citet{kutz:proctor:brunton:2018} or~\citet{mauroy:2021,nakao:mezic:2020} for a particular focus on spectral approximation. 
In view of control,~\citet{arbabi:korda:mezic:2018,peitz:klus:2019,peitz:klus:2020} demonstrate the impact of using the Koopman framework for systems governed by PDEs. 
Moreover, there are first steps regarding nonlinear parabolic problems in one spatial dimension~\citep{deutscher:enderes:2025}.
Symmetry and shift-invariances for PDEs can be leveraged to enable transfer learning and to tackle the high dimensionality usually present in spatial discretizations~\citep{peitz:harder:nueske:philipp:schaller:worthmann:2025}.

\begin{summarybox}
    \textbf{Trading nonlinearity and infinite dimensionality using the Koopman operator}. 
    
    The Koopman operator serves as a widely used tool to obtain linear representations of nonlinear systems by lifting the dynamics onto an infinite-dimensional function space of observables. 
    This linearity enables spectral analysis by means of eigenfunctions and eigenvalues of the Koopman operator, revealing key patterns of the underlying nonlinear systems, such as metastable sets or characteristic modes. 
    As a second key feature of this linearity, it renders the Koopman operator accessible to regression-based approximation such as EDMD, which is accompanied by a comprehensive error analysis using sophisticated tools from ergodic theory, statistical learning, and approximation theory. 
    In particular, recently obtained pointwise error bounds achieved by kernel-based variants lay the foundation for rigorous control-theoretic guarantees such as set-point stabilization or predictive control.  
\end{summarybox}

\section{Linear EDMD with control}\label{sec:EDMDc}
\noindent
EDMDc is a data-driven method for computing approximate linear representations of nonlinear controlled dynamical systems within the Koopman operator framework~\citep{proctor:brunton:kutz:2016,proctor:brunton:kutz:2018}. 
In the following, we first introduce linear EDMDc (Section~\ref{sec:EDMDc:scheme}) and then discuss notable extensions of the basic scheme (Section~\ref{sec:EDMDc:extensions}).

\subsection{The linear EDMDc scheme}\label{sec:EDMDc:scheme}
\noindent
We consider time-invariant nonlinear \emph{controlled} systems in control-affine form
\begin{subequations}\label{eq:dynamics-nonlinear-control-affine}
    \begin{align}
        \dot{x}(t;\hat{x},u) 
        :=&\ \frac{\mathrm{d}}{\mathrm{d}t} x(t;\hat{x},u) 
        \\
        =&\ f(x(t;\hat{x},u)) + G(x(t;\hat{x},u)) u(t)
    \end{align}
\end{subequations}
with locally Lipschitz continuous maps $f:\mathbb{R}^n \to \mathbb{R}^n$, $G:\mathrm{R}^n \to \mathbb{R}^{n\times m}$ with $G(x) = \begin{pmatrix}
    g_1(x) & \cdots & g_m(x)
\end{pmatrix}$, initial condition $x(0;\hat{x},u) = \hat{x}$, and control function $u \in L^\infty_{\operatorname{loc}}([0,\infty),\mathbb{R}^m)$ with the control constraint $u(t) \in \mathbb{U} \subset \mathbb{R}^m$ for a compact set $\mathbb{U}$ containing the origin in its interior, i.e., $0 \in \operatorname{int}(\mathbb{U})$.
Note that control-affine nonlinear systems can be considered essentially without loss of generality, since any general nonlinear control system can be transformed into a control-affine form via a suitable state augmentation and input transformation~\citep[cf.][]{khalil:2002}. 
This observation significantly broadens the practical applicability of Koopman-based control methods, as it allows them to build upon the dynamics~\eqref{eq:dynamics-nonlinear-control-affine}.

As an extension of EDMD, EDMDc incorporates control inputs to approximate the action of the controlled Koopman operator $\mathcal{K}^{\Delta t,\bar{u}}$ for a constant control function $u(t)\equiv \bar{u}\in\mathbb{R}^m$ on the sampling interval with $t\in[t_k,t_k+\Delta t)$, $t_k\geq 0$. 
Here, given snapshot data $\{x_k,u_k,x(\Delta t;x_k,u_k)\}_{k=1}^d$ is used to estimate matrices $A\in\mathbb{R}^{M \times M}$ and $B \in \mathbb{R}^{M \times m}$ satisfying 
\begin{equation}\label{eq:EDMDc:linear-dynamics}
    \Psi(x(\Delta t;\hat{x},\bar{u})) = (\mathcal{K}^{\Delta t,\bar{u}}\Psi)(\hat{x}) \approx A \Psi(\hat{x}) + B \bar{u},
\end{equation}
where $x(\Delta t;\hat{x},\bar{u})$ denotes the solution at time~$\Delta t$ with initial condition~$\hat{x}$ and constant control function $u(t)\equiv \bar{u}\in\mathbb{R}^m$.
To this end, EDMDc solves the least-squares problem 
\begin{equation*}
    \min_{A \in \mathbb{R}^{M \times M},\,B \in \mathbb{R}^{M \times m}} 
    \| \Psi_Y - A \Psi_X - B U \|_F
\end{equation*}
to approximate the lifted dynamics and input influence simultaneously, where $\Psi_X$ and $\Psi_Y$ are the lifted data matrices analogous to the data matrices for EDMD in~\eqref{eq:Koopman:EDMD:lifted-data-matrices}, i.e.,
\begin{eqnarray*}
    \Psi_X & := & \begin{pmatrix}
        \Psi(x_1) & \cdots & \Psi(x_d)
    \end{pmatrix},
    \\
    \Psi_Y & := & \begin{pmatrix}
        \Psi(x(\Delta t;x_1,u_1)) & \cdots & \Psi(x(\Delta t;x_d,u_d))
    \end{pmatrix},    
\end{eqnarray*}
and $U := \begin{pmatrix} u_1 & \cdots & u_d\end{pmatrix}$.
Note that, rather than collecting snapshot data in the original state and, then, lifting it via the vector of stacked observables~$\Psi$, one can collect data $\{\Psi(x_k),u_k,\Psi(x(\Delta t;x_k,u_k))\}_{k=1}^d$ directly in the lifted state. 
This is particularly advantageous if the vector of observables corresponds to sensor outputs, or if not the entire state is measurable.
When truncating to finite-dimensional dictionaries and finitely many samples, the \emph{linear} lifted representation of EDMDc may result in a fundamental approximation error.
More precisely,~\citet{iacob:toth:schoukens:2024} uncover an intrinsic structure of Koopman-based representations of nonlinear systems with inputs, where an accurate lifted representation needs to be at least \emph{bilinear}. 
Interestingly, those limitations also hold true if the input enters linearly into the underlying nonlinear dynamics.

\begin{summarybox}
    \textbf{Linear representations of the Koopman operator for controlled systems might not exist}.
    
    Although the Koopman operator is linear for \emph{autonomous} nonlinear systems, a fundamental limitation in Koopman-based control can already be observed for \emph{controlled} systems with linear control input. 
    Consider the nonlinear dynamics $\dot{x}(t) = f(x(t)) + B u(t)$, where the input~$u$ enters linearly through a matrix $B \in \mathbb{R}^{n\times m}$. 
    Now, assume a \emph{perfect} observable function $\Psi(x)$ based on eigenfunctions of the Koopman operator to find an equivalent (possibly infinite-dimensional) representation of the nonlinear dynamics $\dot{x}(t) = f(x(t))$.
    More precisely, we investigate the corresponding Koopman \emph{generator}, which preserves the control-affine structure~\citep{surana:2016}.
    Then, calculating the Lie derivative of $\Psi$, i.e., its time derivative along the flow of the dynamics, leads to 
    \begin{equation*}
        \frac{\mathrm{d}}{\mathrm{d}t} \Psi(x(t)) 
        = \frac{\partial \Psi(x(t))}{\partial x} f(x(t)) + \frac{\partial \Psi(x(t))}{\partial x} B u(t),
    \end{equation*}
    i.e., the control input may enter the Koopman representation via a state-dependent term for any nonlinear lifting function~$\Psi$, depending on the interplay of~$\nabla \Psi$ and the input matrix~$B$.
    Hence, a linear representation can be achieved if the lifting~$\Psi$ is linear in each component where the input enters. 
    Otherwise, a more sophisticated analysis is required; see, e.g.,~\citet{korda:mezic:2018a}. 
    This reasoning nicely explains the comparatively small regions of attraction for Koopman-based feedback-linearization approaches.
    
    In conclusion, a \emph{linear} finite-dimensional representation \emph{cannot} accurately capture the Koopman dynamics corresponding to a \emph{nonlinear} controlled system in general, even in the simplified case with linear control input in the original system.
\end{summarybox}

Although theoretically limited, linear EDMDc shows remarkable behavior in practical applications, especially when combined with linear quadratic regulation (LQR) or MPC~\citep{korda:mezic:2018a}.
Here, the linear lifted representation enables the use of well-studied prediction and controller design methods from linear systems theory to predict and control general complex nonlinear systems. 
For instance,~\citet{kamenar:crnjaric-zic:haggerty:zelenika:hawkes:mezic:2020} use a linear EDMDc-based lifted model to predict the behavior of a pneumatic soft robot, whereas~\citet{haggerty:banks:kamenar:cao:curtis:mazic:hawkes:2023} employ a similar model with an LQR controller to model and control a dynamic soft robot to achieve real-time reference tracking.
Further, EDMDc can be used to learn and control the dynamics of unmanned aerial vehicles~\citep{narayanan:tellez-castro:sutavani:vaidya:2023,tan:xue:guo:li:cao:chen:2025}.
A detailed collection of the successful application of Koopman operator theory to robot learning can be found in the recent overview article by~\citet{shi:haseli:mamakoukas:bruder:abraham:murphey:cortes:karydis:2024}.
However, control based on linear EDMDc does not come with any closed-loop guarantees and may fail in practice; cf.~\citet{strasser:schaller:worthmann:berberich:allgower:2025,strasser:schaller:worthmann:berberich:allgower:2024b} and Section~\ref{sec:bilinear:EDMDc} for further advantages of bilinear surrogate models as shown, e.g., in~\citet{bruder:fu:gillespie:remy:vasudevan:2021}.

\subsection{Extensions of linear EDMDc}\label{sec:EDMDc:extensions}
\noindent
The finite-dimensional linear EDMDc surrogate model~\eqref{eq:EDMDc:linear-dynamics} is only an approximation, which is inaccurate in general.
More precisely, as with EDMD for uncontrolled systems, there is no a priori guarantee that the space of chosen observables is closed, which generally leads to errors in the representation.
With this motivation, various attempts have been proposed to retain the linear structure of EDMDc while improving the accuracy of the approximation for controller design.

First, the selection of the finite-dimensional basis plays a crucial role in the approximation quality.
As it is known for the autonomous Koopman operator, the associated eigenfunctions of the Koopman operator span an invariant subspace. 
To make use of this fact also for controlled systems, the concept of eigenfunctions may be extended.
One option is to employ a two-step procedure, where first, the eigenfunctions for the autonomous dynamics are learned, and second, after lifting the system to a higher-dimensional linear space, the (linear) input matrix is obtained via regression~\citep{korda:mezic:2020,folkestad:pastor:mezic:mohr:fonoberova:burdick:2020}.
Alternatively,~\citet{kaiser:kutz:brunton:2021} propose to concatenate the state and input space and define an eigenfunction for control similar to an (autonomous) eigenfunction on the extended space.
This allows their utility to be leveraged for (optimal) control in so-called Koopman Reduced Order Nonlinear Identification and Control (KRONIC).
Notably, the authors deduce a Riccati equation with a state-dependent input term which was further explored in~\citet{gibson:calvisi:yee:2022} in the presence of complex eigenvalues.
Nonlinear feedback stabilization is successfully employed in~\citet{huang:ma:vaidya:2020} via control Lyapunov functions and Koopman eigenfunctions. 
In the recent work~\citet{vaidya:2025}, a connection of Koopman theory with Hamilton-Jacobi-Bellman equations is provided. 
Therein, a link between eigenfunctions of the autonomous system and eigenfunctions of extremal dynamics deduced from the Pontryagin maximum principle is given.
In general, applying eigenfunctions when mapping to the higher-dimensional space brings the observable space closer to invariance under the system dynamics and, hence, the resulting approximation error is smaller. 
However, as there is yet no complete error analysis including bounds on the full approximation error, it is not possible to deduce robust closed-loop guarantees when designing controllers based on the linear lifted system.

\citet{asada:solano-castellanos:2024} follow a different path and introduce a physics-based approach called \emph{control-coherent} Koopman modeling.
Here, the dynamical system is split into two subsystems. 
The actuated subsystem is \emph{linearly} driven by the control input $u$ in a direct fashion, while the remaining subsystem is indirectly driven by $u$ through the actuated subsystem.
As a consequence, the lifted system is built based on observables, including the actuated state variables, where the (lifted) input matrix only affects the actuated states. 
Since this structured input matrix exactly mimics the linear input dependence of the original (actuated) state, the input matrix is not approximated and, thus, accuracy is increased.
However, the linear Koopman representation of the unactuated state variables is only exact in infinite dimensions.
Thus, an approximation error occurs when truncating to a finite-dimensional subspace.

The data-driven feedback linearization approach proposed by~\citet{gadginmath:krishnan:pasqualetti:2024} approximates nonlinear system dynamics in a lifted observable space using the Koopman generator.
This representation allows the design of control laws leading to linear closed-loop systems directly from data, illustrating a practical connection between operator-theoretic methods and feedback linearization.
However, the approach relies on derivative measurements and requires an invariant finite-dimensional dictionary for the approximation. 

Besides modifying the lifted linear dynamics, another option is to learn the approximation error.
To this end, \citet{eyuboglu:powell:karimi:2024} estimate a lower bound on the worst-case error, which is then incorporated into a robust frequency-domain controller design.
This approach is refined in~\citet{eyuboglu:karimi:2024}, where the authors learn a suitable integral quadratic constraint (IQC)-based characterization of the residual error of the linear Koopman representation.
The derived error characterization aims to capture both the projection error due to the finite-dimensional dictionary as well as the estimation error due to a finite number of data samples. 
This IQC formulation can then be incorporated into a frequency-based controller design, which is robust w.r.t.\ the residual error to design a controller with closed-loop stability and performance guarantees.
However, the approach relies on a given pre-stabilizing controller if the underlying nonlinear system is unstable. 
Further, the corresponding guarantees are only valid under the assumption that the data-inferred IQC captures indeed the true residual error, which can generally not be validated and contradicts the findings in~\citet{iacob:toth:schoukens:2024}.
\begin{summarybox}
    \textbf{Linear EDMDc is successfully used in practice but lacks closed-loop guarantees}.
    
    Linear EDMDc extends EDMD to control systems by learning a model which is linear in the vector of observables $\Psi(x)$ and in the input $u$.
    This approach has become very popular in recent years and has been successfully applied to various practical control tasks.
    However, the resulting linear surrogate model is, in general, \emph{not} an accurate representation of the underlying nonlinear system.
    Hence, EDMDc with LQR or MPC does not provide any closed-loop guarantees for the underlying nonlinear system and may fail in practice.
    Instead, \emph{bilinear} models are the correct system class that is required to obtain rigorous guarantees for the approximation accuracy and, hence, for the controlled nonlinear system.
\end{summarybox}

\section{Bilinear EDMD with control: Methods and finite-data error bounds}\label{sec:bilinear:EDMDc}
\noindent
As discussed in the previous section, linear Koopman surrogate dynamics are, in general, not sufficient to accurately capture the dynamics of nonlinear control systems.
In this section, we discuss EDMDc schemes with bilinear surrogate models, which are amenable to a rigorous error analysis and therefore pave the way for Koopman-based control with closed-loop guarantees.
In Section~\ref{sec:bilinear:EDMDc:basic}, we discuss the basic bilinear EDMDc scheme, which is more accurate than linear EDMDc, but does not admit guaranteed error bounds.
We then proceed in Section~\ref{sec:bilinear:EDMDc:SafEDMD} by introducing \textbf{S}tability- \textbf{a}nd \textbf{f}eedback-oriented EDMD (SafEDMD), which learns a structured bilinear surrogate model while explicitly bounding the finite-data approximation error.
In Section~\ref{sec:bilinear:EDMDc:kernel}, we discuss kernel-based EDMDc, for which the full approximation error can be quantified.
Finally, Section~\ref{sec:bilinear:EDMDc:discussion} provides a discussion of different error bounds in the context of Koopman-based approximations and their use in control.

\subsection{Bilinear EDMDc}\label{sec:bilinear:EDMDc:basic}
\noindent
In plain bilinear EDMDc, the controlled Koopman action is approximated by a bilinear system
\begin{eqnarray}
    \Psi(x(\Delta t;\hat{x},u)) 
    & = & (\mathcal{K}^{\Delta t,u}\Psi)(\hat{x}) \label{eq:bilinear:dynamics:bilinearEDMDc}
    \\
    & \approx & A \Psi(\hat{x}) + B_0 u + \sum\nolimits_{i=1}^m u_i B_i \Psi(\hat{x}). \nonumber
\end{eqnarray}
The matrices $A\in\mathbb{R}^{M\times M}$, $B_0\in\mathbb{R}^{M\times m}$, and $B_i \in \mathbb{R}^{M \times M}$, $i\in\{1,...,m\}$, are determined based on measured data via the least-squares problem
\begin{equation*}
    \min_{A,B_0,B_1,\ldots,B_m}  \| \Psi_Y - A \Psi_X - B_0 U - \begin{pmatrix}
        B_1 & \cdots & B_m
    \end{pmatrix} \Psi_U \|_F,
\end{equation*}
where $A,B_1,...,B_m \in \mathbb{R}^{M \times M}$, $B_0 \in \mathbb{R}^{M\times m}$, and $\Psi_X$, $\Psi_Y$, $U$ are defined as for linear EDMDc (see Section~\ref{sec:EDMDc:scheme}) and
\begin{equation*}
    \Psi_U := \begin{pmatrix}
        u_1\otimes\Psi(x_1) & \cdots & u_d\otimes\Psi(x_d)
    \end{pmatrix}.
\end{equation*}
As emphasized in the previous section, bilinear dynamics are more accurate representations of nonlinear systems in comparison to linear models; see, e.g.,~\citet{peitz:otto:rowley:2020,bruder:fu:vasudevan:2021,otto:peitz:rowley:2024}.
If there exists a finite-dimensional exact bilinear Koopman representation of a nonlinear system,~\citet{xiong:yuan:miao:wang:cortes:papachristodoulou:2025} propose an extension of Willems' fundamental lemma~\citep{willems:rapisarda:markovsky:demoor:2005} to bypass EDMDc and characterize all possible trajectory via rich enough input-state data.
However, in general, bilinear lifted systems are still only an approximation when restricted to finite-dimensional dictionaries and finite data without imposing strong invariance assumptions; see, e.g.,~\citet{goswami:paley:2021}.
To derive closed-loop guarantees for bilinear Koopman representations, it is thus indispensable to explicitly include the residual error in the controller design and the stability analysis.
To this end, as explained in the following subsection, the data collection and the resulting bilinear surrogate~\eqref{eq:bilinear:dynamics:bilinearEDMDc} need to be slightly modified such that finite-data error bounds can be deduced. 

In~\citet{deutscher:2024}, robust output regulation for LTI systems using a Koopman eigendecomposition obtained via Hankel DMD~\citep{arbabi:mezic:2017b} is presented. 
With the addition of a backstepping controller design, this approach is extended to linear parabolic PDEs on a one-dimensional spatial domain in~\citet{deutscher:zimmer:2025}, building upon the state feedback design presented in~\citet{deutscher:2024}. 
In these works, however, the linearity of the original system allows for the choice of linear observables, resulting in an exact linear representation in the lifted space.
In~\citet{deutscher:enderes:2025}, output regularization of semilinear parabolic systems is achieved via a feedback linearization in the lifted space that eliminates the bilinearity.

\subsection{SafEDMD: Structured bilinear surrogate models with error bounds}\label{sec:bilinear:EDMDc:SafEDMD}
\noindent
In the following, we discuss \emph{SafEDMD}, a bilinear EDMDc variant that combines multiple \emph{autonomous} Koopman operators to recover the controlled dynamics~\citep{strasser:schaller:worthmann:berberich:allgower:2025,strasser:schaller:worthmann:berberich:allgower:2024b}.

To this end, $m+1$ data sets of length $d$, i.e., $\mathcal{D}^{\bar{u}}=\{x_j^{\bar{u}},y_j^{\bar{u}}\}_{j=1}^d$ for $y_j^{\bar{u}} = x(\Delta t;x_j^{\bar{u}},\bar{u})$, mimic $m+1$ \emph{autonomous} versions of the controlled dynamics for a fixed and constant control input $u(t)\equiv \bar{u} \in\{0,e_1,...,e_m\}$, respectively.
Here, the control inputs $e_1,...,e_m$ denote the canonical unit vectors, but any other basis of $\mathbb{U}\subset\mathbb{R}^m$ can be chosen as well.
We emphasize that splitting up data collected for arbitrary inputs may be used to generate suitable synthetic data mimicking the exemplarily chosen ones for $m+1$~fixed input values, analogously to Step~1) of the algorithm presented in~\citet{bold:philipp:schaller:worthmann:2024} and detailed in~\citet{schmitz:bold:philipp:rosenfelder:eberhard:ebel:worthmann:2025}, resulting in flexible sampling. 
Moreover, data efficiency might be further improved by using the approach proposed in~\citet{bevanda:driessen:iacob:toth:sosnowski:hirche:2024}, where future work should investigate how suitable error bounds for control can be obtained within that setting.
Each dataset $\mathcal{D}^{\bar{u}}$ enables the estimation of the autonomous Koopman operator $\mathcal{K}^{\Delta t,\bar{u}}$, respectively.
The finite-dimensional dictionary of observables~$\mathbb{V}$ is chosen as the span of the vector of observables $\widehat{\Psi}:\mathbb{R}^n \to \mathbb{R}^{M+1}$ defined by $\widehat{\Psi}(x) = (1,\Psi(x))$ with
\begin{equation*}
    \Psi = \begin{pmatrix}
        x^\top & \psi_{n+2} & \cdots & \psi_M
    \end{pmatrix}^\top.
\end{equation*}
The observables $\psi_k$, $k\in\{n+2,...,M\}$, are chosen such that $\psi_k\in\mathcal{C}^1(\mathbb{R}^n,\mathbb{R})$ with $\psi_k(0)=0$.
Then, the vector of observables satisfies 
\begin{equation*}
    \|x\|_2 \leq \|\Psi(x)\|_2 \leq L \|x\|_2
\end{equation*}
pointwise for all $x\in\mathbb{X}$ with some $L\in\mathbb{R}$; see~\citet{strasser:schaller:worthmann:berberich:allgower:2025}.
We emphasize that those observable functions can either be inferred from the system dynamics or prior knowledge~\citep{haseli:cortes:2023c,shi:karydis:2021}, or learned via, e.g., a neural network~\citep{takeishi:kawahara:yairi:2017,yeung:kundu:hodas:2019}.
We note that non-state-inclusive dictionaries may be sufficient as long as the state coordinates can be reconstructed (e.g., under suitable observability conditions). 
In system identification, a classical concept to recover the state in the absence of state measurements relies on delay coordinates (see, for example,~\citealp{ljung:1998}), which also show promising results when used as Koopman observables~\citep{mezic:banaszuk:2004,susuki:mezic:2015,brunton:brunton:proctor:kaiser:kutz:2017,korda:mezic:2018a,otto:peitz:rowley:2024}.

To estimate the action of the controlled Koopman operator, the $m+1$ autonomous Koopman operators are estimated from data and control-affinely combined to arrive at the data-driven estimate of the Koopman operator 
\begin{equation}\label{eq:bilinear:Koopman-operator:control-affine}
    \mathcal{K}^{\Delta t,u}_d 
    = \mathcal{K}^{\Delta t,0}_d 
    + \sum_{i=1}^m u_i (\mathcal{K}^{\Delta t,e_i}_d - \mathcal{K}^{\Delta t,0}_d).
\end{equation}
Note that this control-affine structure holds only approximately, as outlined, e.g., in~\citet{peitz:otto:rowley:2020} and rigorously shown in~\citet{philipp:schaller:worthmann:peitz:nuske:2025}.
Since the first observable $\hat{\psi}_0(x(t))\equiv 1$ is constant over time, the Koopman operator and its data-based estimation satisfy a particular structure, where the dynamics of the constant observable are known a priori. 
Thus, this (trivial) dynamics can be removed, while the respective observable results in a state-independent input matrix~$B_0$. 
In particular, this leads to the lifted state~$\Psi$ with dynamics 
\begin{equation}\label{eq:EDMD:dynamics-lifted}
    \Psi(x^+) = A \Psi(x) + B_0 u + \sum\nolimits_{i=1}^m u_i B_i\Psi(x) + r(x,u)
\end{equation}
with residual~$r$ and, using the notation $B_0 = \begin{pmatrix} b_1 & \cdots & b_m\end{pmatrix}$,
\begin{eqnarray*}
        A & = & \argmin_{A\in\mathbb{R}^{M \times M}} \|\Psi_Y^0 - A \Psi_X^0 \|_F \\
        (b_i\ \ B_i) & = & \argmin_{b_i \in \mathbb{R}^{M},\,B_i \in \mathbb{R}^{M\times M}} \|\Psi_Y^{e_i} - (b_i\ \ A + B_i)\, \widehat{\Psi}_X^{e_i}\|_F,
    \end{eqnarray*}
$i\in\{1,...,m\}$, resulting from a linear regression based on the data $\mathcal{D}^{\bar{u}}$, where
\begin{eqnarray*}
    \Psi_X^0 & := & \begin{pmatrix}
        \Psi(x_1^0) & \cdots & \Psi(x_d^0)
    \end{pmatrix},
    \\
    \widehat{\Psi}_X^{e_i} & := & \begin{pmatrix}
        \widehat{\Psi}(x_1^{e_i}) & \cdots & \widehat{\Psi}(x_d^{e_i})
    \end{pmatrix},
    \\
    \Psi_Y^{\bar{u}} & := & \begin{pmatrix}
        \Psi(x(\Delta t;x_1^{\bar{u}},\bar{u})) & \cdots & \Psi(x(\Delta t;x_d^{\bar{u}},\bar{u}))
    \end{pmatrix}
\end{eqnarray*}
for $\bar{u}\in\{0,e_1,...,e_m\}$; see~\citet{strasser:schaller:worthmann:berberich:allgower:2024b} for details.

The idea of interpolating autonomous Koopman operators to define the controlled Koopman operator is not new and has already been exploited, e.g., in~\citet{williams:hemati:dawson:kevrekidis:rowley:2016,surana:2016,peitz:otto:rowley:2020}.
However, SafEDMD is the first framework that also incorporates respective error bounds of the Koopman approximants to build a basis for Koopman-based control with theoretical guarantees.
In particular, it is crucial to establish a \emph{proportional} error bound that vanishes in the controlled equilibrium, e.g., the origin, such that robust control techniques can be used to guarantee closed-loop stability and performance properties.
Importantly, this residual error is due to two sources of error. 
First, the \emph{projection} error $\mathcal{K}^{\Delta t,u} - P_{\mathbb{V}} \mathcal{K}^{\Delta t,u}|_{\mathbb{V}}$ comes from the restriction to a \emph{finite}-dimensional dictionary. 
Second, the enforced control-affine structure in~\eqref{eq:bilinear:Koopman-operator:control-affine} as well as the regression based on a \emph{finite} number of data yields the \emph{learning} error $P_{\mathbb{V}} \mathcal{K}^{\Delta t,u}|_{\mathbb{V}} - \mathcal{K}^{\Delta t,u}_d$. 
For the latter,~\citet[Theorem 3.1]{strasser:schaller:worthmann:berberich:allgower:2024b} establishes a probabilistic point-wise \emph{proportional} error bound, i.e., 
\begin{equation*}
    \|(P_{\mathbb{V}} \mathcal{K}^{\Delta t,u}|_{\mathbb{V}})\widehat{\Psi}(x) - \mathcal{K}^{\Delta t,u}_d\widehat{\Psi}(x)\|_2
    \leq \bar{c}_x \|\Psi(x)\|_2 + \bar{c}_u \|u\|_2 
\end{equation*}
with probability $1-\delta$ for $\delta \in (0,1)$ and $\bar{c}_x,\bar{c}_u\in\mathcal{O}(\nicefrac{1}{\sqrt{\delta d}} + \Delta t^2)$; see~\citet[Proposition 5]{strasser:schaller:worthmann:berberich:allgower:2025} for an analogous result in continuous time.
If a similar proportional bound is available for the projection error with constants $\tilde{c}_x$, $\tilde{c}_u$, the overall residual error is proportionally bounded by 
\begin{equation*}
    \|r(x,u)\|_2 \leq c_x \|\Psi(x)\|_2 + c_u \|u\|_2
\end{equation*}
for each state-control pair $(x,u)\in \mathbb{X}\times\mathbb{U}$ with $c_x=\tilde{c}_x+\bar{c}_x$ and $c_u=\tilde{c}_u + \bar{c}_u$. 
A bound on the projection error trivially holds if, e.g., the dictionary $\mathbb{V}$ is invariant w.r.t.\ the dynamics and, thus, $P_\mathbb{V} \mathcal{K}^{\Delta t,u}|_\mathbb{V} = \mathcal{K}^{\Delta t,u}|_\mathbb{V}$.
Although the invariance of the dictionary is a common assumption in Koopman-based control~\citep{brunton:brunton:proctor:kutz:2016,goswami:paley:2017,huang:ma:vaidya:2018,mauroy:mezic:susuki:2020,goswami:paley:2020,goswami:paley:2021,schulze:doncevic:mitsos:2022,strasser:schaller:worthmann:berberich:allgower:2025}, dynamical systems hardly satisfy this property.
Hence, the projection error should be explicitly characterized to ensure closed-loop guarantees for the underlying nonlinear system.
To address this issue,~\citet{yadav:mauroy:2025,kohne:philipp:schaller:schiela:worthmann:2025} establish uniform error bounds for \emph{autonomous} systems which have been extended to controlled systems in~\citet{bold:philipp:schaller:worthmann:2024,strasser:schaller:berberich:worthmann:allgower:2025} using kernel-based dictionaries.
The basic idea and associated error bounds are explained in the following subsection. 
Moreover,~\citet{guo:korda:kevrekidis:li:2025} propose an approach for combining the learning of such a parameter-dependent Koopman surrogate (the constant control functions~$u$ serve as a parameter) and dictionary learning to further reduce ramifications of the projection step.

\subsection{Kernel-based EDMD with error bounds}\label{sec:bilinear:EDMDc:kernel}
\noindent
As explained in Section~\ref{sec:Koopman:kernel}, kEDMD allows for the formulation of deterministic pointwise bounds on the full approximation error of the autonomous Koopman operator in an RKHS~$\mathbb{H}$.
In the following, we discuss how this analysis can be leveraged to derive proportional bounds on the full residual error for bilinear Koopman surrogates for controlled systems. 
Hence, all trajectories of the underlying nonlinear system are also guaranteed to satisfy the (perturbed) bilinear dynamics in the lifted space for some residual error characterized by the proportional error bound.
In Section~\ref{sec:controller-design}, it will then be explained how this fact can be exploited for controller design.

The kernel-based surrogate model relies on a set $\mathcal{X}$ of $d\in\mathbb{N}$ distinct data points that include the origin and are sufficiently dense in $\mathbb{X}$, i.e., $0\in\mathcal{X}$ and the fill distance is small enough such that at least one kernel feature is non-zero for all $x\in\mathbb{X}$.
For each data point in $\mathcal{X}$, (at least) $m+1$ data triplets $\{x_j,u_{j,l},x(\Delta t;x_j,u_{j,l})\}_{l=1}^{m+1}$ are collected by applying different \emph{constant} control inputs $u_{j,l}\in\mathbb{U}$. 
Here, the applied inputs need to be sufficiently exciting to infer the data-driven surrogate model; see~\citet[Assumption~1]{strasser:schaller:berberich:worthmann:allgower:2025} for details and~\citet{schmitz:bold:philipp:rosenfelder:eberhard:ebel:worthmann:2025} for an in-depth analysis of the excitation condition including an interpretation in terms of subspace angles and optimality conditions.
As outlined in~\citet{schimperna:worthmann:schaller:bold:magni:2025}, the data collection may be significantly weakened to first collect in, e.g., a trajectory-based manner, and then cluster the data afterwards around $d$ artificially introduced cluster centers. 
Another option for less restrictive data collection schemes is based on the kernel-based approach in~\citet{bevanda:driessen:iacob:toth:sosnowski:hirche:2024}.

Similar to SafEDMD, the kEDMD-based estimate of the controlled Koopman operator proposed in~\citet{strasser:schaller:berberich:worthmann:allgower:2025} is constructed as a control-affine combination of autonomous kEDMD estimates. 
The first step is to estimate the nonlinear dynamics via the collected data.
In particular, at each data point $x_j\in\mathcal{X}$, the linear regression problem 
\begin{multline}
    \argmin_{H_j\in\mathbb{R}^{n\times m+1}} \Big\|
        \begin{pmatrix}
            x(\Delta t;x_j,u_{j,1}) & \cdots & x(\Delta t;x_j,u_{j,m+1})
        \end{pmatrix}
        \\
        - H_j \begin{pmatrix}
            1 & \cdots & 1 \\ 
            u_{j,1} & \cdots & u_{j,m+1}
        \end{pmatrix}
    \Big\|_F
    = \hat{H}_j 
\end{multline}
yields the estimates $\hat{H}_j = \begin{pmatrix}
    \hat{f}(x_j) & \hat{G}(x_j)
\end{pmatrix}$ for each $j\in\{1,...,d\}$.
Here, $\hat{f}$ and $\hat{G}$ are estimates of the forward Euler approximations 
\begin{eqnarray*}
    f_\mathrm{fE}(\hat{x}) & = & \hat{x} + \Delta t f(x(t;\hat{x},u)), \\
    G_\mathrm{fE}(\hat{x}) & = & \begin{pmatrix}
            g_\mathrm{fE,1}(\hat{x}) - f_\mathrm{fE}(\hat{x})
            & \cdots & 
            g_\mathrm{fE,m}(\hat{x}) - f_\mathrm{fE}(\hat{x})
        \end{pmatrix}, \\
        g_{\mathrm{fE},i}(\hat{x}) & = & \hat{x} + \Delta t \left( f(x(t;\hat{x},u)) + g_i(x(t;\hat{x},u)) \right).
\end{eqnarray*}
Exploiting the kernel interpolation~\eqref{eq:bilinear:propagation-observable} as outlined in Section~\ref{sec:Koopman:kernel} for each term $f_\mathrm{fE}$ and $g_{\mathrm{fE},i}$, $i\in\{1,...,m\}$, of the approximate dynamics 
\begin{equation*}
    x^+ \approx f_\mathrm{fE}(x) + \sum\nolimits_{i=1}^m u_i (g_{\mathrm{fE},i}(x) - f_\mathrm{fE}(x))
\end{equation*}
yields the (nonlinear) propagation model
\begin{equation*}
    \psi(x^+) \approx \sum_{j=1}^d
        \left(
            \Big(
                \hat{K}_{f_\mathrm{fE}}
                + \sum_{i=1}^m u_i ( \hat{K}_{g_{\mathrm{fE}},i} - \hat{K}_{f_\mathrm{fE}} )
            \Big) \psi_{\mathrm{X}}
        \right)_{\hspace*{-1mm}j} 
    \phi_{x_j}(x),
\end{equation*}
where the estimated function values from $\hat{H}_j$ are used to build $\hat{K}_{f_\mathrm{fE}}$ and $\hat{K}_{g_{\mathrm{fE}},i}$, $i\in\{1,...,m\}$.
\citet{bold:philipp:schaller:worthmann:2024} use this nonlinear surrogate and derive a corresponding approximation error bound when choosing the observable function $\psi(x) = x$. 
Although this enables the usage in, e.g., nonlinear MPC~\citep{schimperna:worthmann:schaller:bold:magni:2025}, the Koopman-based surrogate in the lifted space is still described by \emph{nonlinear} dynamics, which may be challenging for control. 
Building on the results in~\citet{bold:philipp:schaller:worthmann:2024}, the work by~\citet{strasser:schaller:berberich:worthmann:allgower:2025} addresses this issue by proposing a \emph{bilinear} kernel-based surrogate model.
More precisely, a bilinear control-affine kEDMD-based predictor can be established via the approximate control-affine dynamics~\eqref{eq:bilinear:propagation-observable} with a nonlinear data-dependent observable function $\psi$, where~\citet[Theorem 5]{strasser:schaller:berberich:worthmann:allgower:2025} proves a proportional error bound on the full residual error of the kEDMD predictor.
To this end, the authors define the lifted and shifted vector of observables $\Psi(x) = \Phi(x) - \Phi(0)$, where 
\begin{equation*}
    \Phi(x) = \begin{pmatrix}
        \phi_{x_1}(x) & \cdots & \phi_{x_d}(x)
    \end{pmatrix}^\top
\end{equation*}
contains the canonical features corresponding to $\mathcal{X}$.
Note that, using the continuity of $\Psi$, $\Psi(0)=0$, assuming point symmetry of~$\mathcal{X}$ close to the origin, and suitably located data points on the unit sphere, the estimate
\begin{equation*}
    \|\Psi(x)\|_2 \leq L \|x\|_2 \qquad\forall\,x \in \mathbb{X}
\end{equation*}
with some $L>0$ can be inferred exploiting structural properties of the Wendland kernels; see~\citet[Chapter~9]{wendland:2004}. 
Then, the Koopman-based bilinear surrogate dynamics read
\begin{equation}\label{eq:EDMD:dynamics-lifted:kernel}
    \Psi(x^+) = A \Psi(x) + B_0 u + \sum_{i=1}^m u_i B_i \Psi(x) + r(x,u).
\end{equation}
This precisely resembles the dynamics~\eqref{eq:EDMD:dynamics-lifted}, however w.r.t.\ a different dictionary and, thus, also different matrices, i.e., $A=K_{f_{\mathrm{fE}}(\mathcal{X})}^\top K_{\mathcal{X}}^{-1}$, $B_i = (K_{g_{\mathrm{fE},i}(\mathcal{X})} - K_{f_\mathrm{fE}(\mathcal{X})})^\top K_\mathcal{X}^{-1}$, and 
\begin{equation*}
    B_0 = \begin{pmatrix}
        B_1 \Phi(0) & \cdots & B_m \Phi(0)
    \end{pmatrix}.
\end{equation*}
Here, the residual $r(x,u)$ satisfies for all $(x,u)\in\mathbb{X}\times\mathbb{U}$ the error bound
\begin{equation}\label{eq:kEDMDc_error_bound}
    \|r(x,u)\|_2 \leq c_x \|\Psi(x)\|_2 + c_u \|u\|_2
\end{equation}
with $c_x,c_u\in\mathcal{O}(\nicefrac{1}{\sqrt[2N]{d}} + \Delta t^2)$; see~\citet[Theorem~5]{strasser:schaller:berberich:worthmann:allgower:2025} for details.
Here, $c_x$ and $c_u$ approach zero for $\Delta t\to 0$ and $d\to\infty$, where $\Delta t$ needs to converge to zero with a sufficiently fast rate to compensate for the $d$-dependent term $\|K_\mathcal{X}^{-1}\|$.
The error bound in~\eqref{eq:kEDMDc_error_bound} relies on the estimate $\ell \|x\|_2\leq \|\Psi(x)\|_2$ for $\ell>0$.
More precisely, the compact support of the Wendland kernels and the compactness of $\mathbb{X}$ ensure that, outside of a small neighborhood of the origin, the lifted state is bounded away from zero.
Near the origin, the polynomial structure and nonzero directional derivatives of the kernels in all directions guarantee linear growth of the lifting.
Combining these two observations gives a uniform constant $\ell>0$ such that $\|\Psi(x)\|_2 \geq \ell \|x\|_2$ for all $x\in\mathbb{X}$.

\subsection{Discussion}\label{sec:bilinear:EDMDc:discussion}
\noindent
Both the SafEDMD-based surrogate in Section~\ref{sec:bilinear:EDMDc:SafEDMD} as well as the kernel-based EDMD surrogate in Section~\ref{sec:bilinear:EDMDc:kernel} have the same structure. 
In particular, both surrogates yield the bilinear surrogate representation 
\begin{equation*}
    \Psi(x^+) = A \Psi(x) + B_0 u + \sum_{i=1}^m u_i B_i\Psi(x) + r(x,u) 
\end{equation*}
with lifted state $\Psi(x)\in\mathbb{R}^M$ and a proportional error bound on the residual~$r(x,u)\in\mathbb{R}^M$, i.e., 
\begin{equation}\label{eq:proportional-error-bound-discussion}
    \|r(x,u)\|_2 \leq c_x \|\Psi(x)\|_2 + c_u \|u\|_2.
\end{equation}
Hence, designing a controller based on this structure allows for the canonical integration of different Koopman-based surrogate models.
This enables the generalization to novel (and potentially tighter) error bounds in the future, provided that the obtained structure is preserved.

We emphasize that all of the different surrogate learning methods boil down to bilinear system identification.
A vast amount of literature exists on this matter. 
For example, there is literature on identification based on recursive identification~\citep{fnaiech:ljung:1987}, equivalent linear models~\citep{berkhizir:phan:betti:longman:2012}, and bilinear subspace identification~\citep{favoreel:demoor:vanoverschee:2002,verdult:verhaegen:2005,wingerden:verhaegen:2009}.
For the latter,~\citet{favoreel:demoor:vanoverschee:2002} extend the N4SID algorithm~\citep{overschee:demoor:1994} to bilinear systems.
Notably, N4SID identifies the bilinear model by using singular-value decomposition on the data.
In particular, the algorithm estimates the corresponding dimension of the bilinear system's state, unlike EDMD-based methods, which require the definition of a specific lifted state for a chosen lifting dimension beforehand.
However, selecting the state dimension using the singular value threshold in N4SID may result in underfitting or overfitting, which is particularly concerning for high-dimensional systems and noisy data.
Further,~\citet{dasgupta:shrivastava:krenzer:1989} study persistent excitation, and~\citet{sontag:wang:megretski:2009} investigate input selection for identifiability in bilinear systems.
However, these bilinear system identification methods lack suitable (if any) error bounds for control. 
Here,~\citet{sattar:oymak:ozay:2022,sattar:jedra:fazel:dean:2025} prove first finite-sample identification error bounds, however, relying on potentially restrictive assumptions. 
Those bounds have been refined by~\citet{chatzikiriakos:strasser:allgower:iannelli:2024}, establishing data-dependent error bounds that are shown to be significantly tighter for bilinear systems using finite stochastic data.
Notably, the latter also shows initial results for the extension to certain nonlinear systems through the Koopman operator.

Although the established proportional error bounds from SafEDMD and kernel-based EDMD have the same structure, they have different properties.
While SafEDMD allows the use of user-chosen lifting functions $\Psi$, a rigorous error bound on the projection error is missing. 
In addition, the established proportional bound on the learning error holds only with high probability, where the constants $\bar{c}_x$ and $\bar{c}_u$ may be conservative in general.
In contrast, the kernel-based approach yields a deterministic proportional bound on the full residual error stemming from the learning \emph{and} the projection error.
However, the approach relies on data-dependent kernel-based lifting functions whose dimension scales with the data size and, thus, may be impractical for real-world systems.
Therefore, it is crucial to derive the Koopman-based surrogate depending on the use case.
Although the derived error bounds may be hard to compute in practice for general (unknown) nonlinear systems, they offer an explicit dependence on system and data parameters. 
Thus, those error bounds offer qualitative insights and guidelines for the estimation and identification of Koopman-based surrogate models for control.
Here, we note that the particular algorithm used for computing the error bounds is not decisive for establishing closed-loop guarantees.
Rather, it is the underlying structural properties that enable such guarantees.
Therefore, any method that yields \emph{proportional} error bounds for the bilinear surrogate model, as in~\eqref{eq:proportional-error-bound-discussion}, can be employed for a controller design with closed-loop guarantees.

To reduce the conservatism of the established error characterizations in the literature, it may be helpful to use data-dependent error bounds as, e.g., used in~\citet{chatzikiriakos:strasser:allgower:iannelli:2024}.
Interesting future work includes the investigation if suitable (proportional) error bounds for control can also be derived for other modeling approaches, e.g., based on DMDc with Liouville operators~\citep{rosenfeld:kamalapurkar:2024}, the Koopman control family (KCF) framework~\citep{haseli:cortes:2023d,haseli:mezic:cortes:2025} based on the notion of invariance proximity~\citep{haseli:cortes:2023b}, or other learning schemes for (controlled) nonlinear dynamical systems~\citep{sattar:oymak:2022}.
Further, we emphasize that~\citet{iacob:toth:schoukens:2025} propose an \emph{exact} finite-dimensional Koopman embedding for special block-oriented polynomial systems, showing promising results for prediction, and its extension to controller design would be an interesting path for future research. 
Moreover, although~\citet{iacob:szecsi:mate:beintema:schoukens:toth:2025} propose an approach to learn Koopman models under noise by assuming an invariant finite-dimensional dictionary, which is typically non-existent, it remains an open challenge to include noisy output measurements in rigorous error bounds for Koopman-based control.
So far, mostly extensions adding regularization to the regression problem were conducted; see, e.g.,~\citet[Section~2]{bold:philipp:schaller:worthmann:2024}, where the techniques originally developed in~\citet{maddalena:scharnhorst:jones:2021} were leveraged and used in predictive control. 
Another approach is using particularly \textit{exciting} inputs to improve robustness properties in the process of generating the data-driven surrogate model.

\pagebreak
\begin{summarybox}
    \textbf{Bilinear EDMD with control: Proportional error bounds for robust controller design}.
    
    Bilinear EDMD with control provides a data-driven approach to approximate nonlinear dynamics by lifting them into a feature space where state-input interactions take a bilinear form. With suitable construction of the surrogate, one may establish finite-data error bounds on the residual, ensuring rigorous approximation errors enabling closed-loop guarantees. A key property is that the residual error is proportional to the state and input norms, and vanishes at the origin.

    This structure is valuable for control as it does not only provide a global error bound for the data-driven surrogate but also ensures that the dynamics near the equilibrium is captured appropriately.
    In this way, the proportional error bound enables robust controller design since the residual can be explicitly incorporated into design goals. Consequently, controllers synthesized on the bilinear surrogate can be equipped with closed-loop guarantees for the true nonlinear system.

    In conclusion, bilinear EDMD with control allows for connecting data-driven modeling and robust control theory, yielding interpretable, finite-data surrogates whose approximation quality is favorable for feedback control design. 
\end{summarybox}

\section{Controller design based on Koopman models with closed-loop guarantees}\label{sec:controller-design}
\noindent
Optimal control and MPC based on bilinear Koopman models have been very popular and successfully applied in, e.g., robotics; see~\citet{goswami:paley:2021,bruder:fu:vasudevan:2021}. 
However, the accuracy of the Koopman model employed is typically not quantified, and, thus, no guarantees can be given for the controlled nonlinear system.
Guarantees on stability, robustness, and other closed-loop properties are, however, crucial when controlling complex, safety-critical nonlinear systems based on data.

In the following, we discuss control schemes based on bilinear Koopman surrogate models with rigorous closed-loop guarantees.
In Section~\ref{sec:bilinear:EDMDc}, we have discussed how to obtain error bounds for bilinear Koopman models.
In particular, using, e.g., SafEDMD or kernel-based EDMDc, one obtains an uncertain bilinear system
\begin{equation}\label{eq:controller_design_bilinear_model}
    \Psi(x^+) = A \Psi(x) + B_0 u + \sum_{i=1}^m u_i B_i \Psi(x) + r(x,u),
\end{equation}
where $A$, $B_0$, $B_i$, $i \in \{1,...,m\}$, are estimated from the data and the error represented by the residual~$r$ admits a proportional error bound~\eqref{eq:proportional-error-bound-discussion}.
In this section, we discuss controller design approaches for this uncertain bilinear system with stability and performance guarantees for the controlled nonlinear system.
We focus on two complementary approaches, which were considered in the literature:
Feedback design (Section~\ref{sec:controller-design-feedback}) and MPC (Section~\ref{sec:controller-design-MPC}).
Although there exist various examples of successful applications of Koopman-based control, the focus of this section is on methods that derive rigorous closed-loop guarantees for the underlying nonlinear system from data.

\subsection{Feedback design}\label{sec:controller-design-feedback}
\noindent
As discussed in Section~\ref{sec:EDMDc}, a controller based on linear EDMDc comes without closed-loop guarantees for the underlying nonlinear system and may fail in practice. 
In order to address this issue, one must account for the approximation error of the estimated model when using Koopman methods for data-driven control of nonlinear systems.
This naturally leads to a robust control problem:
Stabilizing the bilinear system~\eqref{eq:controller_design_bilinear_model} for all uncertainties $r$ satisfying a given proportional error bound~\eqref{eq:proportional-error-bound-discussion}.
Such an approach was pursued by~\citet{sinha:nandanoori:drgona:vrabie:2022}, where a state-feedback controller is designed to stabilize the (nominal) lifted bilinear system.
However, this work does not explicitly account for the residual error incurred by learning the dynamics from finite data and restricting to a finite set of eigenfunctions.
A first attempt to include the residual error of a Koopman-based bilinear surrogate model in a robust controller design for the underlying nonlinear system is taken in~\citet{strasser:berberich:allgower:2023a}. 
Here, the bilinearity of the surrogate model is over-approximated in a user-chosen region of interest to enable the representation as a linear fractional representation (LFR), which is a common uncertainty characterization in robust control~\citep[cf.][]{zhou:doyle:glover:1996}.
\begin{supplementbox}
    \emph{Supplementary material}: \textbf{Linear fractional representations (LFRs) of uncertain dynamical systems}.
    
    Many control problems involving uncertainties can be formulated via an LFR. 
    To this end, the uncertain system is split into a generalized plant $P$ and an uncertainty $\Delta$, corresponding to the following block diagram. 
    \begin{center}
        \begin{tikzpicture}[%
            auto, node distance=1.5cm and 2cm, >=Latex, 
            block/.style={draw, minimum width=1.5cm, minimum height=1cm, align=center}
        ]
        
          \node[block] (Delta) {\Large$\Delta$};
          \node[block, above=0.5cm of Delta] (M) {\Large$P$};
        
          \draw[->] (Delta.east) -- ++(1,0) |- node[right,pos=0.25] {$w$} ($(M.east) + (0,-0.1666)$);
          \draw[->] ($(M.west) + (0,-0.1666)$) -- ++(-1,0) |- node[left,pos=0.25] {$z$} (Delta.west);
        
          \draw[<-] ($(M.east) + (0,0.1666)$) -- ++(1,0) node[right] {$u$};
          \draw[->] ($(M.west) + (0,0.1666)$) -- ++(-1,0) node[left] {$y$};
        
        \end{tikzpicture}
    \end{center}
    The resulting LFR reads
    \begin{align*}
        \begin{pmatrix}
            y \\ z
        \end{pmatrix}
        &= P \begin{pmatrix}
            u \\ w
        \end{pmatrix} 
        = \begin{pmatrix}
            P_{11} & P_{12} \\ P_{21} & P_{22}
        \end{pmatrix} \begin{pmatrix}
            u \\ w
        \end{pmatrix},
        \\
        w &= \Delta z 
    \end{align*}
    for some uncertainty $\Delta$, where we obtain the mapping from $u$ to $y$ as the linear fractional transformation 
    \begin{equation*}
        \mathcal{F}(P,\Delta) = P_{11} + P_{12} \Delta (I-P_{22}\Delta)^{-1} P_{21}.
    \end{equation*}
    Here, $I-P_{22}\Delta$ needs to be invertible for well-posedness.
    An LFR is a general and powerful representation of uncertainty in dynamical systems, which depends \emph{rationally} on the uncertainty.
    Further information can be found in~\citet{zhou:doyle:glover:1996,scherer:2001b}.
\end{supplementbox}
However, the approach has two main limitations: 
First, the controller design is limited to linear feedback controllers in the lifted state, i.e., $\mu(x) = K \Psi(x)$ for some $K\in\mathbb{R}^{m\times M}$, which leads to a conservative design and comparatively small regions of attraction.
Second, the closed-loop guarantees rely on restrictive assumptions regarding the residual error, which are not theoretically validated and are hard to satisfy in practice. 
To address this theoretical gap,~\citet{strasser:schaller:worthmann:berberich:allgower:2025} exploit the gain-scheduling-based controller design proposed in~\citet{strasser:berberich:allgower:2023b} to reduce conversation by improving the feasibility and, thus, region of attraction of the controller design. 
Further, the therein derived closed-loop guarantees directly couple the proportional residual bound~\eqref{eq:proportional-error-bound-discussion} to Koopman theoretic error bounds on the \emph{learning} error derived in~\citet{schaller:worthmann:philipp:peitz:nuske:2023}. 
This yields an SDP in terms of linear matrix inequalities (LMIs) that parameterize a \emph{nonlinear} control law via convex optimization. 
If the SDP is feasible, the resulting controller
\begin{equation*}
    \mu(x) = \big(I-K_w(I_m \otimes \Psi(x))\big)^{-1} K \Psi(x)
\end{equation*}
with $K\in\mathbb{R}^{m\times M}$ and $K_w \in \mathbb{R}^{m\times Mm}$ is guaranteed to be well posed and to exponentially stabilize the underlying nonlinear system in a certain region of attraction, which is characterized by a sublevel set of the corresponding Lyapunov function.
In particular, the controller ensures positive invariance of that level set w.r.t.\ the closed-loop dynamics and exponential stability of the origin for all initial conditions inside this region of attraction.
The controller design in~\citet{strasser:schaller:worthmann:berberich:allgower:2025} relies on a continuous-time bilinear surrogate model and, thus, requires state-derivative measurements, which are typically difficult to obtain in practice. 
As a remedy,~\citet{strasser:schaller:worthmann:berberich:allgower:2024b} establish an LMI-based controller design via a discrete-time bilinear surrogate based on derivative-free state data, which guarantees closed-loop stability of the underlying \emph{continuous-time} nonlinear system. 
Note that both the design schemes in continuous time and in discrete time solve an LMI-feasibility problem scaling in $\mathcal{O}(M^6)$ with lifting dimension $M$.

Since both~\citet{strasser:schaller:worthmann:berberich:allgower:2025,strasser:schaller:worthmann:berberich:allgower:2024b} rely on an over-approximation of the bilinear term in the lifted dynamics, the resulting robust controller is only feasible for small regions of attraction or small residual errors. 
To make use of the bilinear structure of the lifted dynamical system without over-approximation,~\citet{strasser:berberich:allgower:2025} employ sum-of-squares (SOS) optimization techniques to enhance the discrete-time controller design.
\begin{supplementbox}
    \emph{Supplementary material}: \textbf{Sum-of-squares (SOS) optimization to verify nonnegativity of polynomials}.
    
    SOS optimization is a mathematical technique used in control and systems analysis to verify that certain polynomial functions are always nonnegative.
    By representing a polynomial as a sum of squares of simpler polynomials, this approach enables stability and performance guarantees for nonlinear and uncertain systems, utilizing efficient convex optimization methods. 
    SOS optimization helps to design controllers that can handle uncertainty and nonlinear constraints, making it a useful tool in modern robust control.

    A polynomial matrix $P\in\mathbb{R}[x]^{p\times p}$ of degree $2d$ is said to be SOS if there exists a polynomial matrix $T\in\mathbb{R}[x]^{q\times p}$ of degree $d$ such that $P(x)=T(x)^\top T(x)$. 
    If $P$ is SOS, then $P(x)\succeq 0$ for all $x\in\mathbb{R}^n$.
    However, the converse does not always hold in general. 
    Since verifying the nonnegativity of a polynomial is typically challenging, SOS optimization is an interesting relaxation from a computational perspective~\citep{choi:lam:reznick:1994,reznick:2000,parrilo:2000}.
    In particular, verifying that a matrix is SOS reduces to an SDP in terms of an LMI feasibility problem.
    This so-called Gram matrix method relies on the decomposition $P(x) = (z(x)\otimes I_p)^\top \Lambda (z(x)\otimes I_p)$ for a real matrix $\Lambda=\Lambda^\top\succeq 0$ and a polynomial vector $z\in\mathbb{R}[x]^{l(n,d)}$, where $l(n,d) = \binom{n+d}{d}$ and $x\in\mathbb{R}^n$.
    An open challenge of SOS is its scalability, since the computational complexity $\mathcal{O}((p l(n,d))^6)$ grows rapidly with the problem size, making it impractical for high-dimensional systems or problems with many variables.
\end{supplementbox}
More precisely, a control law with rational dependence on the lifted state is obtained by solving an SDP in terms of an SOS program.
The resulting control law reads 
\begin{equation}\label{eq:control-law-SOS}
    \mu(x) = \frac{1}{u_\mathrm{d}(\Psi(x))} K_\mathrm{n}(\Psi(x)) \Psi(x)
\end{equation}
for a polynomial matrix $K_\mathrm{n}\in\mathbb{R}[z]^{m\times M}$ of degree $2\alpha-1$ and a polynomial $u_\mathrm{d}\in\mathbb{R}[z]$ of degree $2\alpha$ for $z\in\mathbb{R}^M$ and some positive $\alpha \in\mathbb{N}$.
As for the LMI-based controller, the SOS-based controller guarantees closed-loop exponential stability of the underlying nonlinear system.
Notably, the SOS-based controller does not rely on any over-approximation of the (perturbed) bilinear lifted dynamics and, thus, holds globally as long as the corresponding proportional error bound holds globally.
As this error bound is, however, only valid on the sampling region, the resulting closed-loop guarantees for the nonlinear system are again characterized by a Lyapunov sublevel set.
To be precise, this Lyapunov sublevel set is maximized within the sampling region to determine the largest feasible subset of the region of attraction.
Thus, the region of attraction can be significantly enlarged while being able to account for larger error margins than in the LMI-based approach. 
A similar SOS-based controller design has been proposed in~\citet{strasser:berberich:schaller:worthmann:allgower:2025} for the continuous-time controller design. 
Here, the control law does not need to have a rational dependence and the controller, $\mu(x) = K(\Psi(x))\Psi(x)$ with $K\in\mathbb{R}[z]^{m\times M}$ of degree $2\alpha -1$ for $z\in\mathbb{R}^M$ and some positive $\alpha\in\mathbb{N}$, is proven to be exponentially stabilizing within a sublevel set of the Lyapunov function, which is again maximized within the sampling region.

\begin{figure*}[tb]
    \begin{summaryboxOnecolumn}
        \textbf{Koopman-based feedback design with closed-loop guarantees for nonlinear systems}.

        Koopman-based control with guarantees involves two main steps. 
        First, data is collected to construct a \emph{bilinear} Koopman model with proportional error bounds, providing an accurate surrogate of the underlying system. 
        Second, this bilinear model is used to design a robust controller, for example, via an SOS program. 
        If the design is feasible, the resulting control law ensures closed-loop guarantees, such as stability, for the nonlinear system. 
        If not, either the controller's flexibility can be increased using higher-degree polynomial variables, or the model uncertainty can be reduced by collecting more data or refining the dictionary. 
        This procedure systematically integrates Koopman-based modeling and control to provide rigorous guarantees for nonlinear dynamical systems, as illustrated in the figure below, which is adapted from~\citet[Fig.~1]{strasser:schaller:worthmann:berberich:allgower:2024b}.
        \begin{center}
            \begin{tikzpicture}[
                node distance=1.7cm,
                >=stealth,
            ]
                \node (dynSys) [rectangle, rounded corners, minimum width=8cm, minimum height=4cm,draw=black, fill=black!20,align=center,very thick]{~\\[-0.35ex]\large\textsc{unknown nonlinear dynamical}\\[1ex]\large\textsc{ system with control}\\[1ex]\large$\dot{x} = f(x) + g(x) u$\\[0.75cm]};
                \node (DDsurrogate) [rectangle, rounded corners, minimum width=8cm, minimum height=4cm,draw=black, fill=black!20,align=center,very thick] at ($(dynSys.south east)+(3.775,-3.65)$) {~\\[1cm]\large\textsc{Koopman-based bilinear}\\[1ex]\large\textsc{surrogate model}\\[1ex]\large$\Psi(x^+) = A \Psi(x) + B_0 u + \sum_{i=1}^m u_i B_i \Psi(x) + r(x,u)$\\[1ex]\large\textsc{with proportional error bound}\\[1ex]\large$\|r(x,u)\|\leq c_x \|\Psi(x)\| + c_u \|u\|$};
                \draw[->,very thick] ($(dynSys.south west)+(0.5,0)$) |- (DDsurrogate.west) node[pos=.75,below,rotate=0,align=center]{\large\textsc{bilinear EDMD}\\[1ex]\large\textsc{with control}};
                \draw[->,very thick] ($(DDsurrogate.north east)-(0.5,0)$) |- (dynSys.east) node[pos=.75,above,rotate=0,align=center] {\large\textsc{controller with}\\[1ex]\large\textsc{closed-loop Guarantees}};
                \node [rectangle, rounded corners, minimum width=4cm, minimum height=4cm,draw=black, fill=white,align=center,anchor=north west] at ($(dynSys.south west)+(2,0.9)$) {
                    \pgfmathsetseed{1}
                    \begin{tikzpicture}
                        \begin{axis}[
                            domain=-1:1,
                            axis lines=none,
                            xtick=\empty, ytick=\empty,
                            clip mode=individual, clip=false,
                            xmin=-0.5, xmax=0.5,
                            ymin=-0.5, ymax=0.5,
                            width=3.5cm, height=3.5cm,
                        ]  
                            \addplot [black!70, only marks, mark=*, samples=500, mark size=0.75]{rand};
                            \node[anchor=south west,minimum width=0cm,minimum height=0cm,above right,fill=black!20,fill opacity=0.9, text opacity=1,sharp corners] at (axis cs:-1,-1) {$\mathbb{X}$};
                            \draw[ultra thick,black,sharp corners] (axis cs:-1,-1) rectangle (axis cs:1,1);
                            \node[black,anchor=center,fill=black!20,minimum width=0cm,minimum height=0cm,fill opacity=0.9, text opacity=1] at (axis cs:0,0){\large\textsc{data samples}};
                        \end{axis}
                    \end{tikzpicture}
                };
                \node [rectangle, rounded corners, minimum width=4cm, minimum height=4cm,draw=black, fill=white,align=center,anchor=south east] at ($(DDsurrogate.north east)-(2,1.1)$) {
                    \begin{tikzpicture}
                        \begin{axis}[scale only axis,
                            axis lines=none,
                            xmin=-20,
                            xmax=20,
                            ymin=-21,
                            ymax=21,
                            width=6.72cm, height=4.56cm,
                            unbounded coords = jump,
                            restrict x to domain =-25:25,
                            restrict y to domain =-40:40,
                        ]
                            \addplot[black!70,fill=black!70,smooth,mark=*] table [x index=4,y index=5] {figures/data-region-of-attraction.dat};
                            \node[black,anchor=center,fill=black!20,minimum width=0cm,minimum height=0cm,fill opacity=0.9, text opacity=1,align=center] at (axis cs:0,4){\large\textsc{region of}\\[1ex]\large\textsc{attraction}};
                            \node[anchor=south west,minimum width=0cm,minimum height=0cm,above right,fill=black!20,fill opacity=0.9, text opacity=1,sharp corners] at (axis cs:-12,-16.5) {$\mathbb{X}$};
                            \draw[ultra thick,black,sharp corners] (axis cs:-12,-16.5) rectangle (axis cs:12,20.8);
                        \end{axis}
                    \end{tikzpicture}
                };
            \end{tikzpicture}
        \end{center}
    \end{summaryboxOnecolumn}  
\end{figure*}

So far, the discussed control schemes only investigate closed-loop stability guarantees. 
However, in practice, it is often required to optimize the controller to achieve optimal performance in closed loop.
To this end,~\citet{strasser:berberich:schaller:worthmann:allgower:2025} extend the SOS-based controller design from \citet{strasser:berberich:allgower:2025} to account for closed-loop quadratic performance specifications, e.g., an $\mathcal{L}_2$-gain bound, for the underlying nonlinear system. 
We emphasize that the approach is framed based on well-known robust control techniques, such that further extensions to additional uncertainties, like, for example, noise, may be possible in the future. 

Future work should be devoted to the incorporation of less conservative error bounds in the controller design.
In particular, numerical investigations show that the guaranteed error bound for the bilinear surrogate approximation holds already for significantly less data than required by the theoretical analysis. 
Hence, the controller could perform better in closed loop when this gap, due to the theoretical conservatism, would be erased or at least decreased. 
One possibility is to directly use the collected data and learn, e.g., an IQC characterization of the observed residual error.
More precisely,~\citet{eyuboglu:strasser:allgower:karimi:2025} propose to learn a suitable IQC multiplier and parametrization for the residual error seen in the data by transforming the data to the frequency domain. 
Then, this frequency-domain characterization can be used for a frequency-domain-based controller design for performance optimization in, e.g., tracking tasks.
Here, the bilinearity of the lifted dynamics is again over-approximated by an additional IQC to obtain a linear input-output representation of the underlying nonlinear system.
Although the established error bounds tend to be significantly tighter and, thus, improve the closed-loop performance, the bounds are typically only valid in the infinite-data limit. 
Hence, the underlying nonlinear system may not respect the optimized performance bound and, thus, the desired closed-loop guarantees. 
It remains an interesting topic for future work to investigate how the conservatism of a Koopman-based controller design may be reduced while conserving rigorous closed-loop guarantees.

\subsection{Model predictive control}\label{sec:controller-design-MPC}
\noindent
Due to the success of state-feedback controllers based on Koopman models, more advanced control strategies have been combined with the Koopman approach to derive data-driven MPC schemes for nonlinear systems.
In particular, after bringing together (linear) Koopman surrogate models with MPC and showcasing the successful combination of both realms in~\citet{korda:mezic:2018a} (see also~\citet{cibulka:korda:hanis:tomas:2025} for an extension including dictionary learning in the EDMD scheme), several works refined the Koopman-based MPC to investigate theoretical guarantees of Koopman-based MPC and account for bilinear surrogate models.
The recent review paper~\citet{li:yan:zhang:han:law:yin:2025} focuses primarily on linear EDMD and provides an overview of various application areas in Section~$5$. 

In~\citet{arbabi:korda:mezic:2018}, the linear Koopman-based MPC scheme is extended to nonlinear PDEs.
\citet{zhang:pan:scattolini:yu:xu:2022} propose a robust tube-based MPC scheme based on a linear Koopman model with an unknown but bounded approximation error, where, however, the bounded sets are not specified or constructed.
The Koopman-based MPC scheme proposed in~\citet{mamakoukas:dicairano:vinod:2022} relies on a linear surrogate model as well, where the approximation error is assumed to be Lipschitz continuous.
Although this property may be satisfied in any finite region, the true constant is typically not available and, if estimated, leads to a conservative and impractical error characterization for a rigorous closed-loop analysis.
To account for the plant-model mismatch of linear-in-control Koopman models,~\citet{jong:breschi:schoukens:lazar:2024} propose to use a linear MPC scheme with interpolated initial conditions, which yields input-to-state stability of the nonlinear system with respect to the prediction error.

When it comes to \emph{bilinear} Koopman-based prediction models,~\citet{peitz:otto:rowley:2020} illustrate how to use interpolated Koopman generator approximations to formulate a data-driven MPC scheme.
Similarly,~\citet{narasingam:son:kwon:2023} propose a Lyapunov-based MPC scheme using a bilinear Koopman model.
However, these approaches neglect the approximation errors and only validate the closed-loop behavior of the underlying true system numerically. 
Hence, a theoretical investigation is missing.

Assuming invariance of the finite-dimensional Koopman dictionary,~\citet{bold:grune:schaller:worthmann:2025} establish practical stability guarantees for a bilinear EDMDc-based data-driven MPC scheme without terminal ingredients, whereas adding terminal conditions to the MPC scheme yields asymptotic stability of the nonlinear system~\citep{worthmann:strasser:schaller:berberich:allgower:2024}.
Here,~\citet{worthmann:strasser:schaller:berberich:allgower:2024} use a SafEDMD-based state-feedback controller for the terminal control law.
Since invariance of the dictionary is typically hard to satisfy, as for state-feedback controllers, kernel-based EDMD proves useful to capture the full residual error. 
In particular,~\citet{schimperna:worthmann:schaller:bold:magni:2025} use Wendland kernels~\citep{wendland:2004} to define a low-dimensional (nonlinear) Koopman-based predictor using $\Psi(x)=x$ with proportionally bounded approximation error used within data-driven MPC.
Thereby, they establish asymptotic stability of the underlying nonlinear system.
The key properties of the data-driven surrogate model to show that, e.g., cost controllability~\citep{boccia:grune:worthmann:2014,conon:grune:worthmann:2020}, originally proposed in~\citet{grimm:messina:tuna:teel:2005} and further elaborated in~\citet{grune:pannek:seehafer:worthmann:2010}, holds, are the proportional error bound~\eqref{eq:proportional-error-bound-discussion} and a uniform Lipschitz bound.
Notably, both can be directly verified using the bilinear kernel EDMD Koopman control surrogate of Section~\ref{sec:bilinear:EDMDc}.
Further,~\citet{chen:cen:chen:xie:gui:2025} develop a data-driven min–max robust MPC scheme that uses a Koopman-based linear parameter-varying surrogate model with polytopic uncertainty to approximate nonlinear dynamics and guarantees stability and feasibility via LMI constraints.

An interesting future research direction includes the investigation of Koopman-based MPC using the \emph{bilinear} kernel-based surrogate model derived in Section~\ref{sec:bilinear:EDMDc:kernel}.
To avoid solving computationally challenging nonlinear MPC problems, it may be useful to build on results for bilinear data-driven MPC; see, e.g.,~\citet{yuan:cortes:2022,xie:berberich:strasser:allgower:2025}.

\section{Conclusions and future research challenges}\label{sec:conclusion}
\noindent
This work has surveyed recent advances in Koopman-based control methods for nonlinear systems, with an emphasis on frameworks that provide closed-loop guarantees.
While promising results have demonstrated the potential of Koopman-based lifting techniques for analysis, prediction, and control synthesis, the field remains in its early stages, and several fundamental research challenges must be addressed to fully realize robust and scalable Koopman-based control.

\begin{summarybox}
    \textbf{Future research challenges}.
    
    As outlined throughout the paper, we identify the following open research challenges for future work within the community.
    
    \textbf{1) Input-output data}:~
    The existing controller design approaches with closed-loop guarantees rely on input–state trajectories, i.e., assuming that full state measurements are available. 
    In many practical applications, however, only input-output data can be measured. 
    Extending guarantees in Koopman-based control methods to handle this more realistic and restrictive setting remains an open problem. 
    It raises questions about identifiability, observability, and the feasibility of constructing operator-based models that retain control-relevant guarantees when only output data is accessible. 

    \textbf{2) Data requirements}:~
    Finite-data error bounds usable for control usually demand large datasets to approximate the operator accurately. 
    Since these bounds are often conservative, Koopman-based methods may appear to need more data than classical system identification, although, in practice, already less data may suffice. 
    Reducing this conservatism is key to making Koopman-based control with guarantees more broadly applicable in real-world settings.
    
    \textbf{3) Choice of the dictionary}:~
    The dictionary fundamentally governs the balance between model expressivity and computational tractability. 
    Despite progress in automated dictionary learning (e.g., via deep learning), there is still no consensus on how to ensure that these learned representations preserve control-relevant features of the underlying nonlinear dynamics and enable the derivation of closed-loop guarantees.
    One interesting question is whether ideas, e.g., from bilinear subspace identification~\citep{favoreel:demoor:vanoverschee:2002,verdult:verhaegen:2005,wingerden:verhaegen:2009} can circumvent the prior selection of the dictionary and allow for implicit definitions through the identification procedure.
    
    \textbf{4) Scalability}:~
    While convex formulations and linear control synthesis in the lifted space offer computational advantages, ensuring that stability and performance guarantees translate back to the original nonlinear dynamics remains difficult. 
    Current approaches often impose strong assumptions -- such as noise-free data -- or introduce conservatism, which limits practical applicability and scalability to high-dimensional or distributed systems. 
    Methods that can reduce conservatism and cope with noisy data without sacrificing closed-loop guarantees represent a crucial direction for future research.
\end{summarybox}

In summary, while Koopman-based approaches hold considerable promise for unifying nonlinear control with linear operator theory, their practical success will hinge on addressing these challenges. 
Future research should focus on principled strategies for input-output data learning, dictionary design, scalable yet tractable controller synthesis with reduced conservatism, and frameworks that provide robust closed-loop guarantees for nonlinear safety-critical systems evolving in dynamically changing environments. 
Addressing these problems will be key to transitioning Koopman-based control from a primarily theoretical tool to a widely adopted methodology in complex real-world applications.


\bibliographystyle{cas-model2-names}

\bibliography{literature}                               

\end{document}

%% file: figures/reprojection.tex
\begin{center}
{\begin{tikzpicture}[scale = 1.0]
    \draw[smooth, tension=0.5] plot coordinates{(1,2) (2,2.4) (4,1.78) (6.3,2.8)}{};
    \draw[smooth, tension=0.5] plot coordinates{(6.3,3.8) (4.5,3) (2,3.5) (1.1,3)}{};
    \draw[smooth, tension=1] plot coordinates{(1,2) (1.1,2.5) (1.1,3)}{};
    \draw[smooth, tension=1] plot coordinates{(6.3,2.8) (6.4,3.4) (6.3,3.8)}{};
    \node at (6.5,4.1) {$\Psi(\mathbb{R}^{n})$};
    \draw[line width=.08em] (0.5,0)--(1,1)--(7,1)--(6.5,0) -- cycle;
    \draw[line width=.08em,->] (0.5,0) -- (0.5, 4) node [label=right:$\mathbb{R}^{M-n}$] {};

    \draw[smooth, tension=0.5] plot coordinates{(2,0.2) (2.5,0.5) (3.5,0.2) (4,0.4)} node[above left=-.05cm] {$x(\Delta t;\hat{x})$};
    \draw[thin, fill=black] (2,0.2) circle (1.5pt) node[label=left:$\hat{x}$]{};
    \draw[thin, fill=black] (4,0.4) circle (1.5pt);

    \draw[smooth, tension=0.5] plot coordinates{(2,2.8) (2.5,2.9) (3.5,2.3) (4,2.5)} node[above=-.01cm] {$\Psi(x(\Delta t;\hat{x}))$};
    \draw[thin, fill=black] (2,2.8) circle (1.5pt) node[label=above:$\Psi(\hat{x})$]{};
    \draw[thin, fill=black] (4,2.5) circle (1.5pt);
    \draw [dotted] (2,2.8) -- (2,0);
    \draw [dotted] (4,0.4) -- (4,2.5);
    
    \draw[thin,blue,fill=blue] (5,1.5) circle (1.5pt);
    \node (a) at (5.55,1.5) {\textcolor{blue}{${K}\Psi(\hat{x})$}};
    \draw [dashed] (4.98,1.55) -- (4.7,2.4);
    \draw[thin, fill=black] (4.7,2.4) circle (1.5pt);
    \node (a) at (7,0.2) {$\mathbb{R}^{n}$};
    \draw [dashed] (4.7,2.4) -- (4.7,0.8);
    \node (c) at (5.5,0.5){$\Psi^{-1}\!\!\ \circ \, \pi_W \circ\, {K}\Psi(\hat{x})$};
    \draw[thin, fill=black] (4.7,0.8) circle (1.5pt);
\end{tikzpicture}}
\end{center}